\documentclass[preprint,12pt]{elsarticle}
\usepackage[utf8]{inputenc}

\usepackage{amssymb}
\usepackage{amsmath}

\usepackage{geometry}
\usepackage{graphicx} 
\usepackage{graphicx} 
\usepackage{tikz}
\usepackage{pifont}
\usepackage[cc]{titlepic}

\usepackage{natbib}
\usepackage[T1]{fontenc}
\usepackage{enumitem}
\usepackage{wrapfig}
\usepackage{amsmath,amssymb,amsfonts}
\usepackage{graphicx}
\usepackage{changepage}
\usepackage{textcomp}
\usepackage{subfigure}
\usepackage{booktabs}
\usepackage{xcolor}
\usepackage{environ}
\usepackage{algorithmicx}
\usepackage{algpseudocode}
\usepackage{algorithm}
\usepackage{textcomp}
\usepackage{amsmath}
\usepackage{amssymb}
\usepackage{stmaryrd}
\usepackage{float}
\usepackage{scalerel}
\usepackage{flexisym}
\usepackage{xcolor, colortbl}
\usepackage{multirow}
\usepackage{hhline}
\usepackage{graphicx}
\usepackage[export]{adjustbox}
\usepackage{graphicx} 
\usepackage{subcaption}
\usepackage{graphicx}
\usepackage{verbatim}

\usepackage{lineno}
\usepackage[nottoc]{tocbibind}
\usepackage[Symbol]{upgreek}
\usepackage{mathtools}
\usepackage[colorlinks=true, urlcolor=blue]{hyperref}
\begin{document}
\title{SEER: \textit{S}ustainability \textit{E}nhanced \textit{E}ngineering of Software \textit{R}equirements
}
\author[1]{Mandira Roy}
\affiliation[1]{organization={Ca' Foscari University},
            city={Venice},
            country={Italy}}

\author[2]{Novarun Deb}
\affiliation[2]{
organization={University of Calgary},
            city={Calgary},
            country={Canada}}
\author[3]{Nabendu Chaki}
\affiliation[3]{
organization={University of Calcutta},
            city={Kolkata},
            country={India}}
\author[1]{Agostino Cortesi}


%
\begin{abstract}
    The rapid expansion of software development has significant environmental, technical, social, and economic impacts. Achieving the United Nations Sustainable Development Goals by 2030 compels developers to adopt sustainable practices. Existing methods mostly offer high-level guidelines, which are time-consuming to implement and rely on team adaptability. Moreover, they focus on design or implementation, while sustainability assessment should start at the requirements engineering phase.
In this paper, we introduce \textit{SEER}, a framework which addresses sustainability concerns in the early software development phase. The framework operates in three stages: (i) it identifies sustainability requirements (SRs) relevant to a specific software product from a general taxonomy; (ii) it evaluates how sustainable system requirements are based on the identified SRs; and (iii) it optimizes system requirements that fail to satisfy any SR. The framework is implemented using the reasoning capabilities of large language models and the agentic RAG (Retrieval Augmented Generation) approach.  \textit{SEER} has been experimented on four software projects from different domains. Results generated using Gemini 2.5 reasoning model demonstrate the effectiveness of the proposed approach in accurately identifying a broad range of sustainability concerns across diverse domains.
   
\end{abstract}

\maketitle
\section{Introduction}
Sustainability is a global concern with multifaceted dimensions that involve environmental, economic, technical, and social perspectives \cite{Seghezzo01072009}. It requires attention from all industries, including the software sector. Rapid growth of the software industry over the past few decades has contributed to significant resource consumption, energy wastage, social inequalities, and a substantial carbon footprint, among other impacts. Many of the 17 Sustainable Development Goals (SDGs) \footnote{https://sdgs.un.org/goals} are directly related to the software development industry. 
Although some large organizations like Meta are adopting sustainable practices, such as renewable energy data centers, these efforts are not consistently seen across small- or medium-scale software enterprises. Sustainable system development should include not only resource optimization, but also how the software is designed and the services it intends to provide.

Sustainable software development faces several key challenges \cite{10.1145/3709352}. These include a lack of systematic approaches for sustainability assessment, difficulties in making trade-off decisions between sustainability criteria, and insufficient methods to correlate sustainability theory with non-functional system properties. Most often, there is a lack of adequate knowledge or a superficial understanding of sustainability principles. Existing research works propose methodologies for sustainable software development, such as organizational behavior changes \cite{9725722}, green computing \cite{10815134}, measuring energy efficiency and the carbon footprint of applications \cite{10791178}. Some other works have focused on the introduction of sustainability in specific phases of software development, such as requirements engineering \cite{9604527, 10260926}, and green coding practices \cite{10734439}.  It is worth noticing that most of these current studies focus mainly on the environmental dimensions of sustainability, while other dimensions are often not well incorporated or studied in isolation \cite{bambazek00}. However, the different dimensions of sustainability are interrelated and should be considered throughout the software development lifecycle \cite{10803467}.

A common challenge is that while system stakeholders may express a desire for sustainable software \cite{SULEMANA2025100445}, these requirements are often not explicitly or granularly stated, unlike clearly defined functional requirements (FR) and non-functional requirements (NFR)\cite{Le2016RelatingAD}. This challenge can be addressed by, identifying the SDGs that are directly or indirectly impacted by that particular system and also managing the conflicts that possibly arise concerning the FRs and NFRs, to keep the specification consistent. The software industry and researchers have mapped the SDGs into definite sustainability requirements (SRs) based on different domains \cite{c84c6ba3995d4529a2537028f16dcea1}. These specific SRs have to be identified and compared with the system specification.


Sustainability requirements are considered as a special category of quality requirements that have a larger objective \cite{10803467}, \cite{9604527}: satisfaction of sustainability requirements is a concern of global stakeholders, who may not always be direct users but are affected by its impact. Some common concerns relevant to a software system that defines sustainability requirements are:
\begin{itemize}
    \item[-] "\textit{Is the software system possibly discriminating users on gender, age, physical ability, etc.?}"
    \item[-] "\textit{What might be the carbon footprint of a large-scale usage of the system?}"
    \item[-] "\textit{Does the system provide sustainable pricing models?}"
    \item[-] "\textit{Does the system design ensure longevity of its operation?}". 
\end{itemize}
The challenge is to identify which of these concerns are relevant for different software products based on their defined scope. For example, sustainable pricing models for end-user is an important SR for educational software. Instead, in the case of banking software an important SR would be the transparency of transactions. Therefore, thes SR identification highly depends on the scope defined for the software product. 

The sustainability concerns are often confused \cite{9218220} with traditional NFRs defined for the system. 
It is important to stress the fact that sustainability requirements address the interests of a broader set of stakeholders than those considered to be the target of the software product, by considering the impact on the environment and on the economic and social context.
 SRs may impose constraints on functionalities so that they are inclusive, reusable, modular and so on. Similarly SRs may define further constraints on NFRs based on standard guidelines \cite{10.1145/2593743.2593744} to check whether security protocols ensure enough privacy, access to information and others.

The primary contribution of this work is the design of the \textit{SEER} framework and the implementation of a supporting tool for SR engineering based on it. In its first component \textit{SEER} identifies the SRs relevant to a given software application within the specific product scope. It then evaluates how existing FRs and NFRs align with and influence these SRs. Finally, it recommends adjustments to FRs and NFRs to ensure a satisfactory alignment with the targeted SRs. 



To the best of our knowledge, this is the first tool introduced in the literature that supports the requirements engineering process by eliciting related sustainable requirements. The proposal has been assessed using publicly accessible requirement datasets. This assessment provides both qualitative and quantitative evidence on how requirements are modified to incorporate sustainability considerations within the system.

The structure of the paper is organized as follows. Section \ref{rel} elaborates on existing state of the art. Section \ref{prelim} introduces some foundational concepts for the reader. Section \ref{seer} illustrates the \textit{SEER} framework. Section \ref{exp} provides experimental results and Section \ref{con} concludes.

\section{Related Works} \label{rel}
This section provides an overview of research works addressing different dimensions of sustainability at the software development process level (or early) rather than only at the product level. This section mainly covers some recent research works on different areas, like surveys showing sustainability scenarios in the software industry, proposals on early inclusion of these concerns and specific works on particular sustainability aspects.

Becker et al. \cite{7202997} have designed the Karlskrona manifesto that provides the principles of sustainability design for software-intensive systems. The manifesto covers three main dimensions of sustainable software design- environmental, social and economic. It proposes an initial set of 8 principles. 

Ribeiro et al. \cite{10.1145/3605098.3636031} have conducted a literature review to understand the current state of sustainability in software engineering. Their findings show that sustainability is still a new topic among software engineers. There are only a few approaches, methods, processes and tools to support sustainable software development. Survey research \cite{su14148596} demonstrates how software professionals perceive sustainability. Their main findings show that 92\% of professionals were not able to identify SRs for software development. The research work has also presented SRs for a word processing and food application. Bambazek et al. \cite{9830111} have also conducted survey to show sustainability impacts within agile software development. They have examined which factors of Scrum \cite{8229928} framework can be used to enhance sustainability. Their results suggests that Scrum is a suitable option for sustainable software development.

Several existing works have highlighted on why it is necessary to consider sustainability concerns at the earliest in software design. Karita et al.  \cite{10.1145/3555228.3555236} have conducted qualitative data analysis to understand which sustainability dimensions are considered within SDLC (Software Development Life Cycle). Their findings indicate that technical, environmental, and social concerns are present in all SDLC phases. Authors have suggested that SRs should be considered in early SDLC phases and software quality requirements are directly related to sustainable development. Gross et al. \cite{10803467} have proposed a conceptual framework to support integration of sustainability from the very beginning in software design. They have considered four main dimensions (social, technical, environmental and economic) and have shown how software quality and software sustainability are related and can be mapped together. König et al. \cite{10.1145/3709352} have shared their vision of how software engineering can contribute to the fundamental sustainability goals by 2030. They have outlined key steps for sustainability-driven software engineering, which include developing a nuanced understanding of sustainability concepts, tool support for sustainability assessment and others.

Seyff et al. \cite{DBLP:conf/re/SeyffBGSCDPVB18} have introduced a crowd-focused platform for evolution of SRs in the requirement engineering phase. The platform enables to systematic elicitation, analysis and negotiation of requirements. Saputri et al. \cite{9604527} have introduced a goal-scenario-based approach to integrate sustainability concerns within the software development in the early requirement engineering phase. They have propose an approach Enhanced Sustainability Requirements Engineering (ENSURE) to elicit stakeholder needs across five sustainability dimensions, analyse risk and priority of goals and resolve conflicts.

Some research works have focussed on particular aspects of sustainability. Moreira et al. \cite{MOREIRA2023102107} have designed a sustainability catalog for social and technical dimensions in ICT systems. Their catalog includes different categories and sub-categories across the dimensions. They have used goal modeling approach to show inter-relations among the categories. Restrepo et al. \cite{doi:10.3233/AO-230004} have proposed an ontology of critical sustainability factors for software development. They have considered five dimensions of sustainability: environmental, technical, economic, social, and individual/personal. They have mapped different software quality attributes to these dimensions. Pham et al. \cite{9218220} have suggested nine different dimensions of sustainability- social, economic, environmental, technical, individual, purpose, design-aesthetics, integrative and legal. They have proposed a conceptual framework, ShapeRE, that helps project stakeholders to reason about sustainability concerns in different project phases. This framework evaluates different artifacts considering the sustainability dimensions. 

Hassan et al. \cite{su16145901} have proposed a machine learning based approach for correlating NFRs and SRs. They have trained BERT \cite{devlin2019bertpretrainingdeepbidirectional} language model using promise\_exp \footnote{https://zenodo.org/records/12805484} dataset. The NFR categories are associated to the sustainability dimensions.

The literature survey on sustainability for software development has led to the following observations: (i) several survey works have highlighted that SRs are important and must be considered within SDLC; (ii) the considerations of SRs must begin with the requirement engineering phase itself; (iii) elicitation of SRs are manual and stakeholder-dependent; (iv) correlation among SRs and system requirements are done at broader (or abstract) level. The first two observations directly contributes to the motivation of our proposed approach. Eliciting SRs solely from stakeholders may be inadequate due to gaps in their knowledge and understanding of sustainability concepts. Therefore, it is essential to also account for global policies and country-specific regulations. Most researchers have linked sustainability dimensions to system requirements in general, this alignment can be achieved at a much more granular level by directly considering explicitly defined sustainability goals and principles.


\section{Preliminaries}\label{prelim}

\subsection{Sustainability Requirements}
The rapid boom in the software development industry calls for some attention towards their impact on different sustainability dimensions \cite{10803467}.  
In this research work, we have considered four main dimensions to which SRs can be mapped- environmental, social, technical and economic.

\begin{figure*}[!h]
            \centering
            \includegraphics[width=0.8\linewidth]{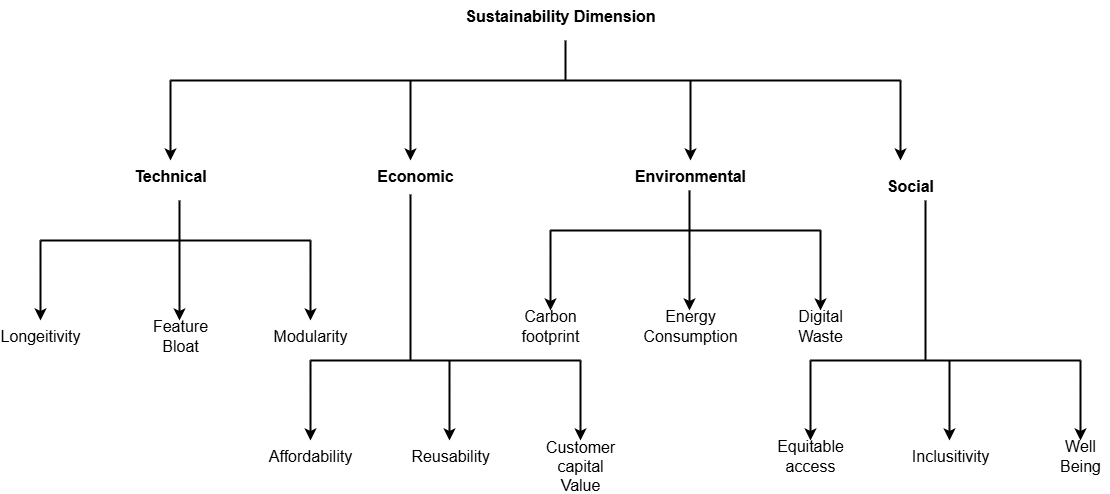}
            \caption{SR Types within each dimension}
            \label{fig1}
        \end{figure*}

Each sustainability dimension has multiple concerns that falls into different categories. For each dimension, we can categorize the SRs into different types based on the specific concern it is related to.  Figure \ref{fig1} shows some categories (broader concern) within each dimension of sustainability. It is to be noted we have considered only those types that can be associated with software design and its functionalities.

\subsection{Sustainability Requirements and NFRs}
The SRs are considered as a special category of NFRs. Unlike traditional NFRs, they are not always directly elicited from system stakeholders but are often motivated by societal and environmental concerns.
Traditional NFRs define quality constraints to improve the security, reliability and performance of the services. However, sustainability NFRs differs from this as they define how the software should operate to minimize negative impacts on environmental, social, technical and economic aspects. Let us consider the scenario of a healthcare application system (refer to Table \ref{eg}).  Such a system have some very specific performance and security constraints (refer to row 2 of Table \ref{eg}). The first constraint is associated with a service that fetches patient records and the second service is associated with a service that stores patient records in the system. However, SRs provide a broader objective rather than each functionality. The last three rows of Table \ref{eg} show the SRs associated with such a system for four different dimensions. 

\begin{table}[!h]
\caption{NFRs and SRs for Healthcare System}
\label{eg}
\begin{tabular}{|l|l|}
\hline
\textbf{Type}          & \textbf{Description}                                                                                                                                                                                                            \\ \hline
\textbf{NFRs}          & \begin{tabular}[c]{@{}l@{}}The system shall return search results for patient records \\ within 2 seconds,\\ The system shall encrypt all patient data using AES-256 \\ encryption during storage and transmission\end{tabular} \\ \hline
\textbf{Environmental} & \begin{tabular}[c]{@{}l@{}}The system shall minimize energy consumption by optimizing \\ server load and supporting green cloud providers\end{tabular}                                                                          \\ \hline
\textbf{Social}        & \begin{tabular}[c]{@{}l@{}}The application shall support multiple languages and accessibility \\ features to accommodate users with disabilities or limited literacy\end{tabular}                                               \\ \hline
\textbf{Economic}      & \begin{tabular}[c]{@{}l@{}}The system shall be designed to minimize licensing and \\ maintenance costs for deployment in low-resource clinics\end{tabular}                                                                      \\ \hline
\textbf{Technical}     & Reducing feature bloat that leads to inefficient software                                                                                                                                                                       \\ \hline
\end{tabular}
\end{table}

The satisfaction of SRs, however is dependent on the stakeholder-defined functionalities and traditional NFRs of the system. Let us consider an FR such as - "\textit{The system shall provide real-time chat between patients and doctors during online consultations}". Such an FR may have a negative correlation with social sustainability concerns as accessibility and multilingual support may delay or distort communications. Thus, it is important to assess how the stakeholder-defined requirements of the system correlate with SRs. 

\subsection{Sustainability Requirements Taxonomy}\label{sectaxo}
\begin{figure}
    \centering
    \includegraphics[width=0.7\textwidth]{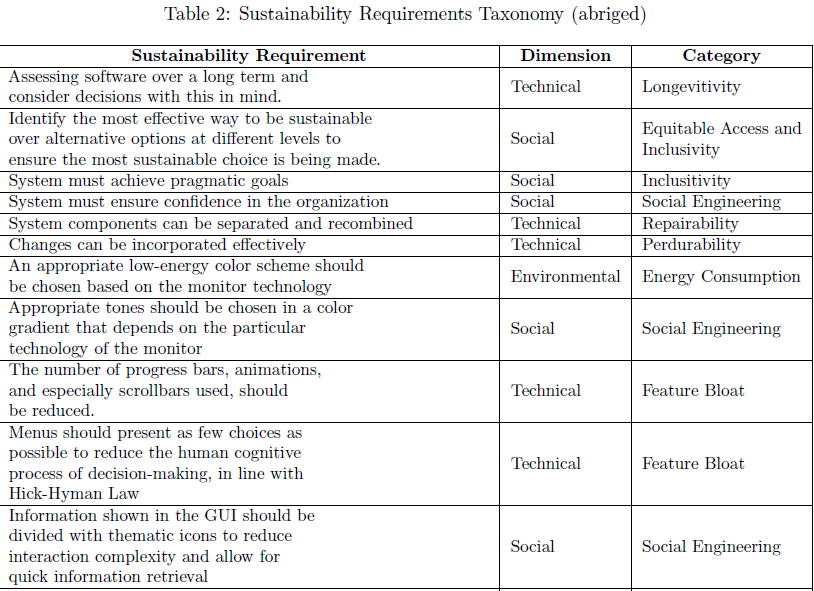}
    \caption{An instance of Taxonomy \cite{ourwork}}
    \label{fig:taxo}
\end{figure}

Several research works in the existing literature have proposed sustainability concerns or requirements for different application domains \cite{9735126}, \cite{10260926}, \cite{ouhbisofia}. Some works even explored particular dimensions of sustainability \cite{ceuralhinai}, \cite{GARCIABERNA2021124262}, \cite{inbookstoumpos}. The proposed \textit{SEER} framework is intended to be applicable for any software application domain, considering four main sustainability dimensions- environmental, economic, social and technical. The implementation of  \textit{SEER} framework depends on the availability of a comprehensive SRs taxonomy. In this research work, we have referred to the taxonomy defined in \cite{ourwork}. The taxonomy presented \cite{ourwork} is generic and contains SRs on all four dimensions mentioned above. The taxonomy is constructed by aggregating data from existing state-of-the-art. It consists of three fields- (i) Sustainability Requirements, (ii) Dimension and (iii) Category. Figure \ref{fig:taxo} shows an instance of the taxonomy.

\section{The \textit{SEER}  Framework}\label{seer}
\begin{figure*}[!h]
    \centering
    \includegraphics[width=0.7\linewidth]{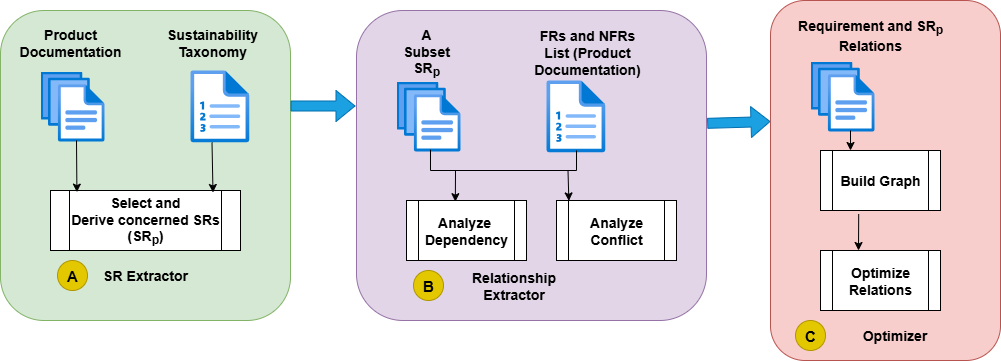}
    \caption{\textit{SEER} Framework}
     \label{fig:framework}
\end{figure*}

The  \textit{SEER} framework aims to integrate sustainability constraints within  software design process. The proposed framework enhances the system's requirements by- (i) introducing SRs relevant to that system, (ii) identifying and managing possible conflicts arising from the resulting requirements set and (iii) fine-tuning requirements to be compliant with sustainability objectives. The framework consists of three main components. Figure \ref{fig:framework} shows the components of the requirement analysis framework. The components are as follows-
\begin{enumerate}
    \item \textit{SR Elicitor}- This component excerpts the most relevant sustainability requirements ($SR_p$) for a software product from the taxonomy (refer to section \ref{sectaxo}).
    \item \textit{Relationship Integrator}- This component identifies the interrelationship among the system functional and quality requirements with the derived $SR_p$.
    \item \textit{Sustainability Optimizer}- This component suggests suitable modifications to system requirements to enhance their satisfiability towards sustainability requirements ($SR_p$). 
\end{enumerate}
The following sections describes each of these components in detail.

\subsection{SR Elicitor Component}
\begin{figure*}[!h]
    \centering
    \includegraphics[width=\textwidth]{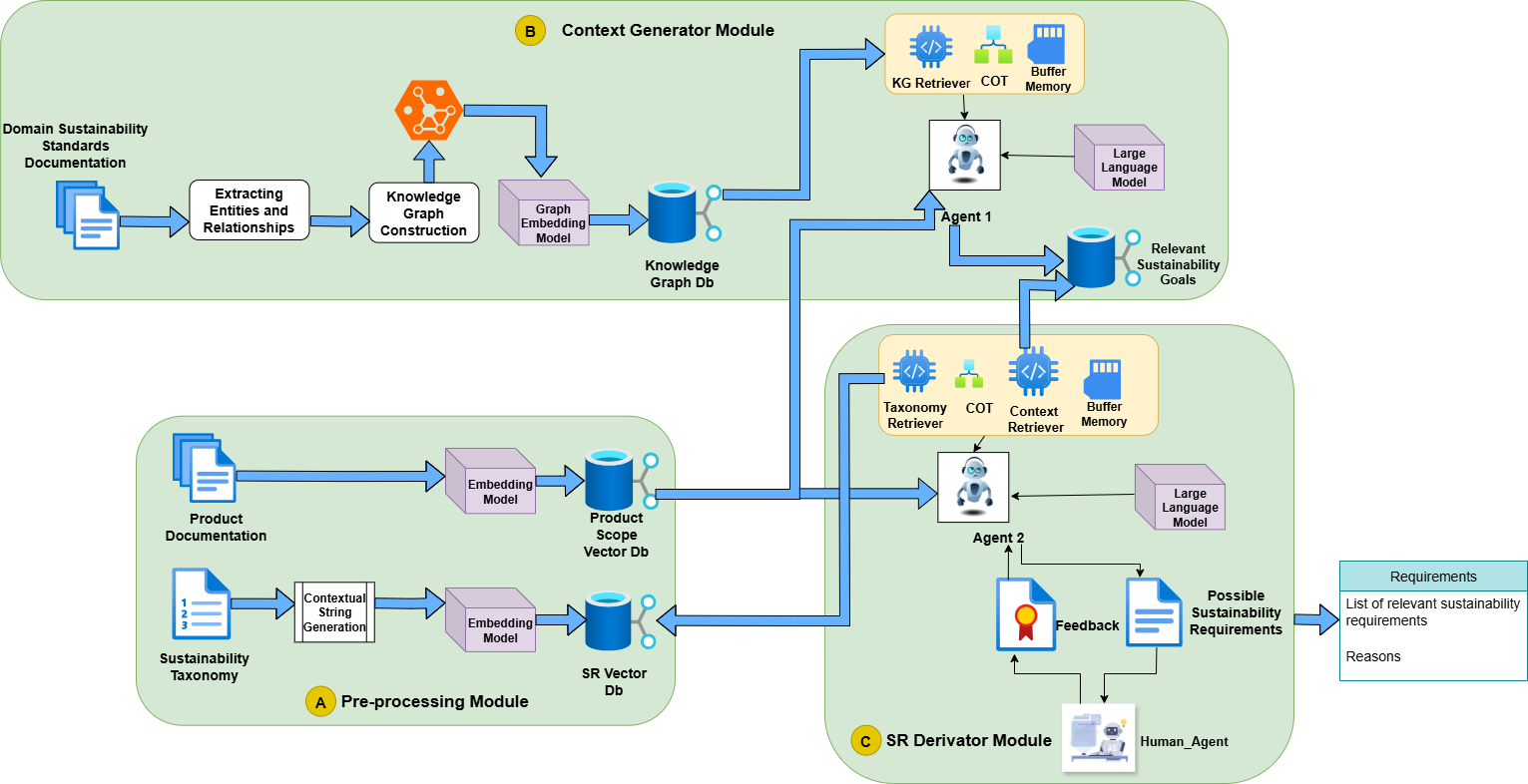}
    \caption{SR Elicitor Component}
    \label{fig:SRExtract}
\end{figure*}

The  \textit{SR Elicitor} component derives relevant SRs from the taxonomy based on the overall product scope, operational and development constraints and other such high-level descriptions of the product. Figure \ref{fig:SRExtract} shows the detailed structure of this component. The component consists of three different modules that operate as follows- 

\subsubsection{Pre-processing Module}
In this module, the input documents- software product documentation and sustainability requirements taxonomy are pre-processed and stored in a vector database for usage in the next modules. 

    \begin{enumerate}
        \item \textit{Pre-processing Product Documentation}- In this phase, we focus only on the general product scope, which covers the overall product description, product features, user classes and characteristics, design and implementation constraints, and the operational environment. Each of these elements are usually documented as separate sections within the product specification document. The system’s FRs, external interface requirements, and NFRs are addressed in the subsequent phase of the framework.

At first, each section of the general product scope is reviewed and refined to eliminate ambiguity using standard tools \cite{10.1145/3540250.3558928}. Next, the semantic consistency across these sections is validated before the document is used. This is done by embedding each section using a sentence transformer model \footnote{\label{sentarns}https://huggingface.co/sentence-transformers/all-mpnet-base-v2}. The cosine similarity between the embedded sections is calculated to measure coherence. If the similarity exceeds a predefined threshold (at least 0.5), the document is considered acceptable; otherwise, it must be revised before proceeding. In a real scenario, the requirement analyst will be informed about non-coherence and must take responsibility for refinement. This is important so that the \textit{SEER} framework correctly interprets the overall product scope.
    
    The pre-processed software product's documentation is then semantically chunked. Each chunk represent each section of the general product scope. Then, text embedding is applied on the chunks using a language model $^{\ref{sentarns}}$ and stored in a vector database (refer to product scope vector Db in Figure \ref{fig:SRExtract}).

    \item \textit{Pre-processing Sustainability Taxonomy}- The data in the sustainability requirements taxonomy is available in tabular format (as shown in Figure \ref{fig:taxo}), consisting of three fields- SR, dimension and category within that particular dimension. The data in these three fields are concatenated to form a contextual string. For example, if we consider the SR in the first row of Table \ref{fig:taxo} the contextual string would be- "The requirement is to \textit{Assessing software over a long term and consider decisions with this in mind} under the dimension \textit{Technical} and falls in the category of \textit{Longeitivity}". The generation of these contextual strings is intended to make each SR complete and meaningful. 
    
    The sustainability taxonomy is divided into chunks, with each contextual string represented as an individual chunk. These chunks are then embedded using the same language model $^{\ref{sentarns}}$ applied to the product documentation. The resulting embeddings are stored in a vector database (see SR Vector DB in Figure \ref{fig:SRExtract}).  
    
\end{enumerate}

\subsubsection{Context Generator Module}
In this module, we utilize a domain-specific sustainability standard document, such as those published by governments (e.g., Canada's Federal Sustainable Development Strategy \footnote{\label{casus}https://www.canada.ca/en/services/environment/conservation/sustainability/federal-sustainable-development-strategy/departmental-strategies.html}, UN's 2030 Agenda for Sustainable Development \footnote{https://sdgs.un.org/2030agenda}), for context embedding. The standard document is transformed into a knowledge graph, which then helps map a product to its relevant SRs.

\begin{figure*}[!h]
    \centering
    \includegraphics[width=0.8\textwidth, frame]{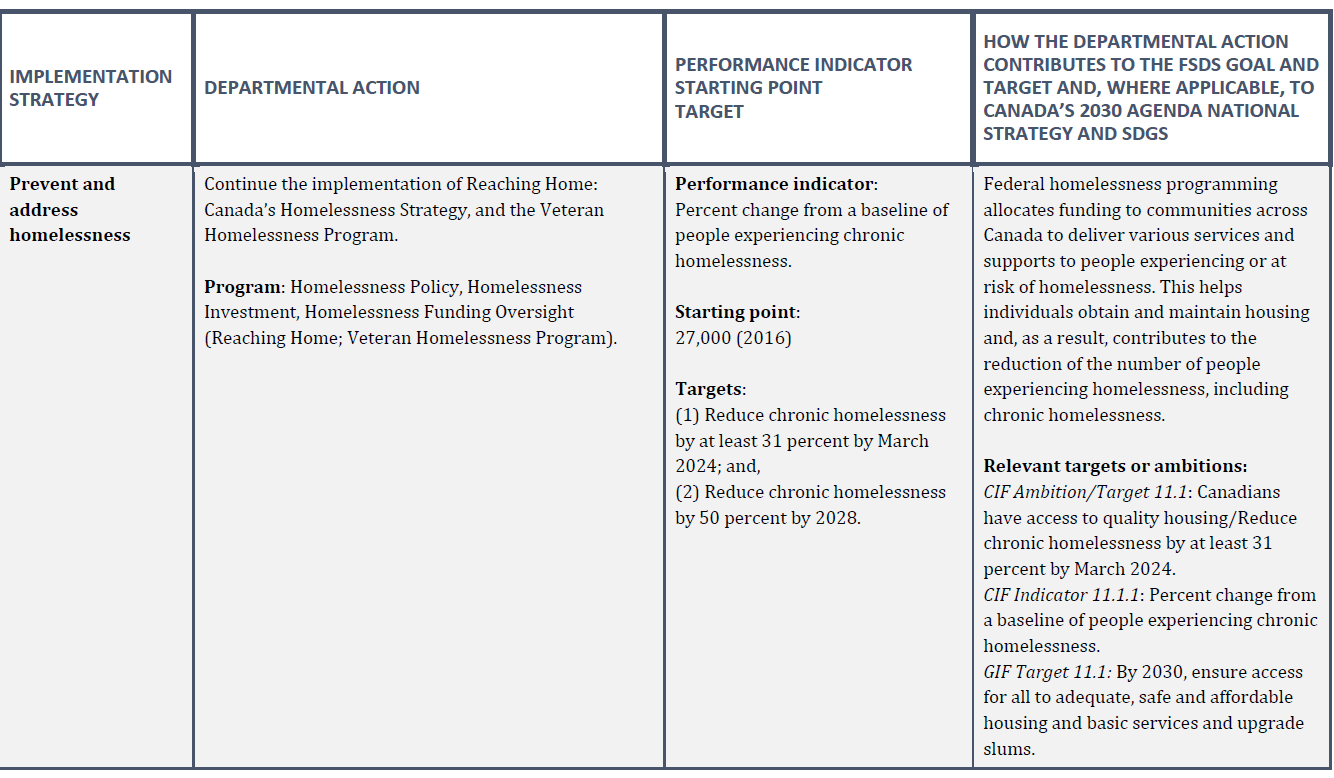}
    \caption{Snapshot of Sustainability Standard Document for Canada Housing}
    \label{fig:standard}
\end{figure*}

These standard documents are issued by government organizations that outline a country's policies and goals for achieving sustainability. They define various sustainability objectives across different domains like healthcare, transport, and housing. When developing a software system to meet sustainability goals, it's essential to consult such a standard document. Figure \ref{fig:standard} shows an instance of the guidelines developed by Canadian government for housing domain.


The \textit{Context Generator} module operates in two steps-

\begin{enumerate}
    \item \textit{Knowledge Graph Construction}- In this step, the standard document is processed and transformed into a knowledge graph. The knowledge graph is build considering an ontology. The details of the ontology are provided in the appendix.
    
    Based on this ontology, different entities (goals, targets, indicators) and relationships have to be identified from the document. These entities and relationships are identified by taking the help of an large language model (LLM) like ChatGPT \cite{ye2023comprehensivecapabilityanalysisgpt3}. The LLM is prompted with a structured one-shot prompt (defined using \footnote{\label{claude}https://docs.claude.com/en/docs/build-with-claude/prompt-engineering/overview}) consisting of (i) the ontology it must follow and (ii) an example knowledge graph built on a small document (refer to Appendix). The results obtained from LLM are stored in a file and passed for human validation. The manual verification process may also involve any required modification. After manual verification, the information is represented in a graph structure. Finally, similar embeddings  (as used for product document chunks $^{\ref{sentarns})}$ are defined for the entities and relations in the graph and stored in a graph database such as neo4j \footnote{\label{neo4j}https://neo4j.com/product/auradb/}. Figure \ref{fig:graph} shows an instance of the knowledge graph.
    
    \begin{figure*}[!h]
    \centering
\includegraphics[width=0.6\textwidth, frame]{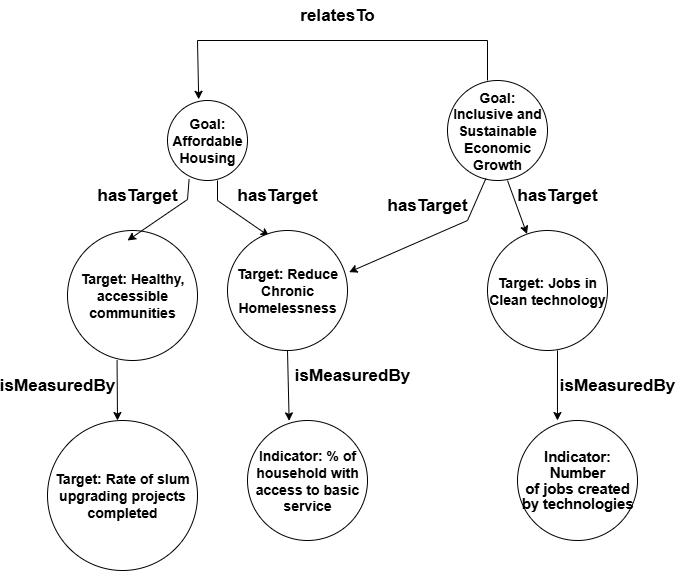}
    \caption{An instance of a Knowledge Graph for Housing standard}
    \label{fig:graph}
\end{figure*}
\item \textit{Knowledge Extraction}- In this step, an agentic Graph RAG approach is applied to derive sustainability goals, targets and indicators from the knowledge graph that must be considered for a software product. The sub-steps of knowledge extraction are as follows-

\begin{itemize}
    \item An LLM acts as the brain of the agent, and it is provided with tools such as- knowledge graph-retriever tool (referred as KG Retriever in Figure \ref{fig:SRExtract}), COT (chain-of-thought) prompt and conversational buffer memory\footnote{\label{memory}https://python.langchain.com/api_reference/langchain/agents/langchain.agents.openai_functions _agent.agent_token_buffer_memory.AgentTokenBufferMemory.html}. The details of the COT prompt (template referred from \footnote{\label{agent}https://python.langchain.com/api_reference/langchain/agents.html}) are provided in the Appendix section of the manuscript.
    \item Each product chunk is fetched from the \textit{product scope vector Db} (refer to Figure \ref{fig:SRExtract}) and provided to the agent.
    \item The agent operates through the following steps- (i) interprets the task using the CoT prompt, (ii) analyzes the provided chunk to understand the product scope, and (iii) employs the KG retriever to extract relevant goals, targets, indicators, and relationships. The agent uses the retrieved information to reason over the results and formulates a final answer. If needed, it may perform multiple retrieval cycles before reaching the final response.
    \item The agent's responses for each chunk are saved in memory.
    \item After all chunks are processed, the agent revisits the memory. It summarizes the entities derived by processing each chunk and provides a final complete list of goals, targets, indicators and relationships.
\end{itemize}

The outcome of this step is a set of sustainability goals, targets, indicators and relationships among them that are relevant to the particular product scope. This extracted context is stored in a vector database to be used as a context for the next module.

\end{enumerate}

\subsubsection{SR Derivator Module}
This module of \textit{SR Elicitor} component finally derives the actual SRs from the taxonomy defined in \cite{ourwork} that must be considered while implementing the software product. The modules operate in the following two steps-
\begin{enumerate}
    \item \textit{SR Listing}- In this step, an agentic RAG approach is applied to extract relevant SRs from the taxonomy corresponding to each chunk of the product documentation. The sub-steps of SR listing are as follows-
    \begin{itemize}
        \item The agentic RAG approach consists of an LLM as the brain of the agent, and tools such as- taxonomy retriever, context retriever, COT prompt and conversational buffer memory.
        \item Each product chunk is fetched from the \textit{product scope Db} (refer to Figure \ref{fig:SRExtract}) and fed to the agent.
        \item The agent at first understands the task through the COT prompt. It then processes the product chunk and uses the context retriever tools to derive relevant sustainability goals and targets. Based on the derived context, it further retrieves semantically related SRs from the taxonomy. The agent reasons over the derived information from the two sources and provides a final answer.
        \item The agent response for each chunk is stored in memory.
        \item The agent then summarizes the SRs derived by processing each product chunk and provides a final, complete list of SRs.
    \end{itemize}
  This final list of SRs are fed into the next step for human validation.  

\item \textit{Expert Analysis}- In this step, a human agent (a domain expert) verifies the list of SRs generated by the agentic RAG approach and gives feedback. If the human expert finds any inconsistency in the result, the feedback is used as input (in the prompt) and the agentic RAG approach is executed again. The process continues until the human expert approves. The outcome is a list of sustainability requirements $SR_p$ and LLM-generated reasoning as to why these should be considered. 
\end{enumerate}

\subsection{Relationship Integrator Component}
\begin{figure*}[!h]
    \centering
    \includegraphics[width=0.6\textwidth]{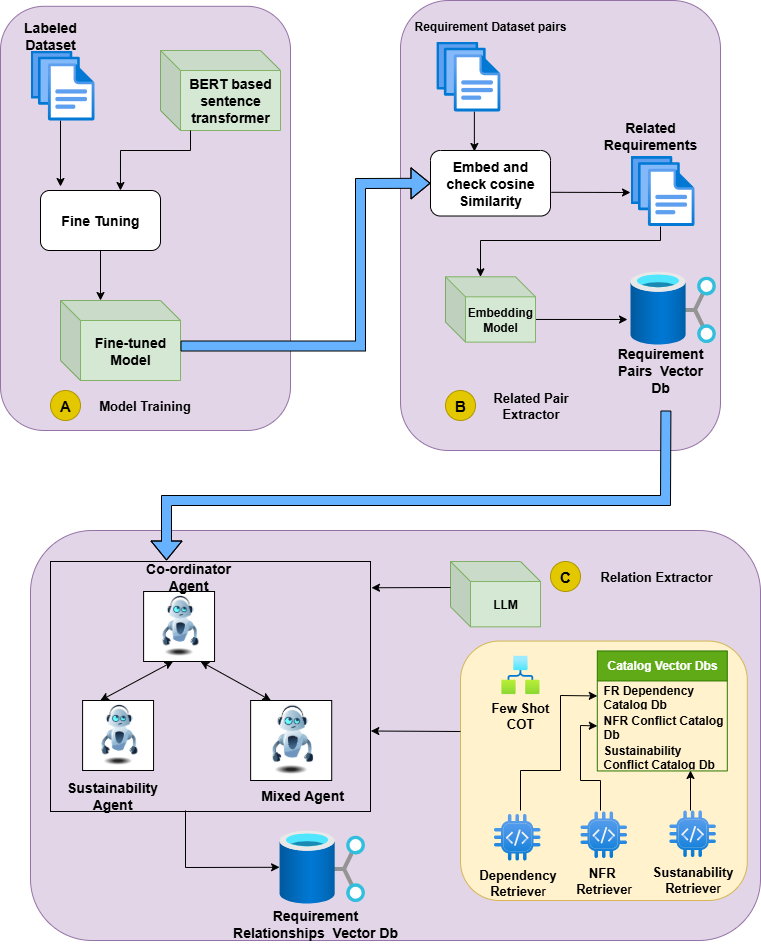}
    \caption{Relationship Integrator Component}
    \label{fig:relextrct}
\end{figure*}

This component analyzes and derives relationships among the derived sustainability requirements ($SR_p$) and the system requirements (functional and non-functional requirements). Figure \ref{fig:relextrct} shows the architecture of the \textit{Relationship Integrator} component. It consists of three modules - \textit{Model Trainer}, \textit{Related Pair Extractor} and \textit{Relation Extractor}. The operation of these three modules are as follows-

\subsubsection{Model Training}
The objective of this module is to fine-tune a sentence transformer model$^{\ref{sentarns}}$ for correlating related requirement pairs. The total set of possible requirement pairs considering $FRs$, $NFRs$ and $SR_p$ may be sufficiently large. So in the initial two modules, we aim to reduce the number of requirement pairs to the ones that are semantically related and then in the third module, we derive their relationship type. The training process consists of the following steps-
    \begin{enumerate}
        \item At first, a labeled dataset of requirements pairs is prepared. This dataset is an aggregation of labeled requirement pairs obtained from the works of \cite{10.1007/978-3-030-93709-6_29}, \cite{9218190}, \cite{su16145901}. In the dataset the requirement pairs that are semantically related are marked as 1, otherwise 0. There total 5000 labeled requirement pairs in the dataset \footnote{\label{github link}https://github.com/MCompRETools/SEER}.
        \item A pre-trained sentence transformer model$^{\ref{sentarns}}$ is used as the base model for generating embeddings of requirement statements.
        \item The sentence transformer model is fine-tuned on the labeled dataset$^{\ref{github link}}$ so that it can better capture requirement-specific semantics.
    \end{enumerate}
The output of this module is the fine-tuned model that specializes in identifying semantically similar requirements. 
\subsubsection{Related Pair Extractor}
The objective of the module is to use the fine-tuned model for identifying pairs of requirements (for the target system) that are semantically related. Before executing the module, candidate pairs of requirements are created. Three different types of pairs are created (using system $FRs$, $NFRs$ and derived $SR_p$). The first type of pair links two different SRs, allowing the identification of potential interdependencies between them. The second type of pair links an FR with an SR to determine whether the FR supports or conflicts with the SR. Similarly, the third type of pair links an NFR with an SR, and it is used to identify how system-level NFRs influence the SRs. In this process, every FR and every NFR is paired with each SR in the set $SR_p$
The steps are as follows-
    \begin{enumerate}
        \item These candidate pairs are at first embedded using the fine-tuned model generated in the previous module. 
        \item The embedded pairs are then checked for cosine similarity. Any requirement pair with a cosine similarity score greater than or equal to 0.65 is marked as related. The threshold is chosen after carefully experimenting with different values multiple times (with experimental datasets) to select only those pairs that hold some relations.
        \item The related requirement pairs are then stored in a vector database so that semantically related requirement pairs can be efficiently retrieved later.
    \end{enumerate}
    The output of this module is a subset of requirement pairs that are semantically related.
 \subsection{Relation Extractor}
 The \textit{Relation Extractor} module derives the exact nature of the relationship (positive or negative) among the related requirement pairs. This module employs a multi-agent-based LLM system \cite{tran2025multiagentcollaborationmechanismssurvey} to identify these relationships. The multi-agent system primarily consists of two classes of agents-
    \begin{itemize}
        \item \textit{Coordinator Agent}- The coordinator controls the flow of information and assigns tasks to specialized agents.
        \item \textit{Specialized Agents}- There are two types of specialized agents.
        \begin{enumerate}


            \item \textit{Sustainability Agent} – This agent specializes in determining relationships (dependency type) among the first type of pairs of requirements (consisting of SRs only).

            \item \textit{Mixed Agent} – This agent specializes in determining relationship types among pairs where two different types of requirements are involved (second and third pair type). 
        \end{enumerate}
    \end{itemize}
    Each of these agents is supported with different COT prompts (based on template $^{\ref{agent}}$), knowledge catalogs and retriever tools. The details of the prompts are provided in the appendix. Three types of knowledge catalogs are created- (i) a catalog of possible types of dependencies among FRs and their definitions, (ii) a catalog of NFR correlations (absolute or relative conflicts), (iii) a catalog of correlations among different sustainability categories within different dimensions (positive and negative influence). The first two catalogs are directly obtained from the sources in the literature \cite{10.1007/978-3-642-23391-3_3}, \cite{9222252}. The third catalog is created by taking the help of GPT \cite{gpt3.5}. GPT was prompted to generate correlations among SR categories. The results generated by GPT are then tuned by the authors. Each author has separately marked the correlations generated by GPT with agree or disagree. The first author then combined the results and conducted a final discussion round (with all authors) to determine which correlations should be considered within the catalog. Each of these catalogs is embedded and stored in a vector database. Since there are three different catalogs, three different retriever tools are defined. The catalogs are available at our GitHub repository $^{\ref{github link}}$. 

    The \textit{Sustainability Agent} has access to the third catalog, which defines dependency types among SRs. The \textit{Mixed Agent} has access to all three catalogs. The purpose of these catalogs is to offer insight into the various types of relationships that can exist between requirements. The LLM agent can consult these catalogs during reasoning to determine whether a particular relationship type applies to a given requirement pair.
    
    The operational steps of this module are as follows-
    \begin{enumerate}
    \item Each requirement pair is retrieved from the vector database.
        \item The \textit{Co-ordinator Agent} at first determines the type of requirement pair.
        \item Based on its type it calls one of the \textit{Specialized Agent}. 
        \item A LLM-based agent understands the task through few-shot COT prompt. It then fetches required knowledge from one or more vector databases (of catalogs). The agent reasons over the retrieved information and derives the relationship type between the requirements. The agent also supports its response with a reasoning.
        \item The \textit{Specialized Agent} returns the final response (relationship type and reason) to the \textit{Co-ordinator Agent}.
        \item The \textit{Co-ordinator Agent} provides a formatted output consisting of the relation and reason determined for each requirement pair.
   \end{enumerate}
The output of \textit{Relation Extractor} is the identified and classified relationships among requirement pairs (dependencies, conflicts, sustainability impacts) with the help of the LLM and agent collaboration. The identified relations are stored in a vector database.

\subsection{Sustainability Optimizer Component}
\begin{figure*}[!h]
    \centering
    \includegraphics[width=0.6\textwidth]{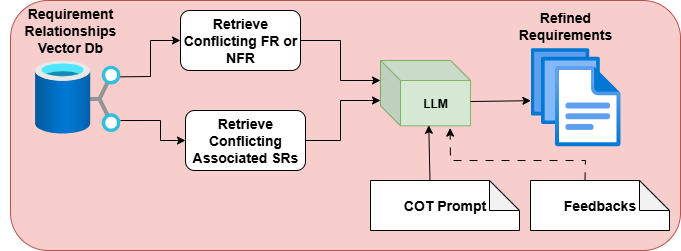}
    \caption{Optimizer Component}
    \label{fig:opcomp}
\end{figure*}

The objective of this component is to refine the FRs and NFRs that conflicts with the required sustainability requirements ($SR_p$). Figure \ref{fig:opcomp} shows the architecture of the \textit{Optimizer} component. The operational steps of this component are as follows-
\begin{enumerate}
    \item The recorded relationship between requirement pairs is embedded and stored in a vector database.
    \item A retriever function is defined that fetches requirement pairs where an FR or NFR has a negative correlation with the corresponding SR.
    \item Let the retrieved pair be $<R_j, SR_k>$. $R_j$ is an FR or NFR and $SR_k$ is the associated SR.
    \item Another retriever function is defined that fetches all other SRs that are in conflict with $SR_k$ and are also associated with $R_j$. 
    \item An LLM client is defined and it is provided with the following inputs-
    \begin{itemize}
        \item Step-wise COT prompt (refer to appendix)
        \item Requirement pair $<R_j, SR_k>$
        \item Associated $SR$ interdependencies retrieved in previous step.
    \end{itemize}
    The conflicting $SR$ relations are also provided to the LLM, so that refined $R_j$ must maintain positive correlation with all associated $SRs$. The LLM client is invoked to generate a revised $R_j$ that positively correlates with the SR $SR_k$.
    \end{enumerate}
  
    The refined FR or NFR has to be validated in a two-step process.
    \begin{itemize}
        \item The refined FR or NFR and SR is then again passed through \textit{Relationship Integrator} component to check if the correlation has changed to positive.
        \item The refined FR or NFR is subjected to a system analyst for their review. This step is important as the refined requirement must fit within the project scope and requires human intervention.
    \end{itemize}
     The outcome of both steps can further be merged and the optimizer module can be executed again, considering the two-step feedback. This system analyst can decide whether to accept the result or execute the optimizer module again.
   
\subsection*{Completeness Checking}
This is an additional step of \textit{SEER} framework that aims to ensure that all derived $SR_p$ are satisfied by at least one system requirement. A simple check algorithm is executed after the optimization is done to check if all $SR_p$ are included in the final list of requirement pairs. If there exists any SRs that are not satisfied by any of the system requirements then, system analyst has to be notified. 

The \textit{SEERTool} available at $^{\ref{github link}}$ is an implementation of the proposed framework. It consists of all the three components and also has the support for completeness checking.

\section{Experimental Evaluation}\label{exp}
The aim of this Section is to assess how effectively the proposed framework supports the elicitation of sustainability requirements across software in various domains. The experimental process analyzes the software requirement specification (SRS) document and elicits sustainability concerns and conflicts with respect to given standards or guidelines. The datasets used in the experimental process are as follows:
\begin{enumerate}
    \item We have selected case studies from the PURE dataset, a well-known  benchmark dataset of software requirement specifications that  has been widely used by researchers in machine learning and NLP (natural language processing) approaches \cite{11008757}, \cite{11088115}. 
    We selected four of case studies on very different application domains, holding high relevance in different sustainability areas:
    smart-home applications highlight concerns regarding energy efficiency and resource management, healthcare applications mainly emphasizes social sustainability; transport systems applications focus on environmental, technical and economic challenges; and e-commerce applications reflect economic, social and environmental dimensions. These diverse domains are chosen to verify the effectivness of \textit{SR Elicitor} in correctly selecting SRs.  Table \ref{tabdata} provides an overview of the selected case studies. 
\begin{table*}[!h]
\caption{Details of selected PURE datasets}
\label{tabdata}
\resizebox{\textwidth}{!}{%
\begin{tabular}{|l|l|c|c|}
\hline
\textbf{Domain}   & \textbf{Case-study Description}                                                                                                                                                & \#\textbf{FRs} & \#\textbf{NFRs} \\ \hline
Smart Home       & \begin{tabular}[c]{@{}l@{}}Specifies the requirements for the development of a \\ “Smart House”, called DigitalHome\end{tabular}                           & 41            & 12             \\ \hline
Healthcare       & \begin{tabular}[c]{@{}l@{}}Describe requirements for the Nenios Child Care \\ Management (NCCM) software\end{tabular}                                      &    27           &   13             \\ \hline
E commerce       & Specifies requirements for GAMMA-J’s Web Store                                                                                                             &      6         &   38             \\ \hline
Transport System & \begin{tabular}[c]{@{}l@{}}Defines the application software requirements for\\ the  Interstate -15 Reversible Lane Control System\\ 
(I-15 RLCS)
\end{tabular} & 35              &            13    \\ \hline
\end{tabular}%
}
\end{table*}
\item The UN SDG document\footnote{https://sdgs.un.org/goals} provides a set of overall sustainability goals that applies for a vast range of domains. However, there are limited number of standard documents that frames the SDGs within specific domain and regional issues. The \textit{SEER} framework can be applied to any standard guidelines. In our experimental evaluation, for building the knowledge graph, we referred to the Canadian sustainability standard guidelines\footnote{https://www.canada.ca/en/services/environment/conservation/sustainability/federal-sustainable-development-strategy/departmental-strategies.html}, one of the most specific, comprehensive and structured documentation we came across. 
These guidelines are openly available and are defined separately for different domains.  The second column of Table \ref{tabstan} lists the references for the documents used in each case study. 

\end{enumerate}

\begin{table*}[!h]
\caption{Knowledge Graph Creation Details}
\label{tabstan}
\resizebox{\textwidth}{!}{%
\begin{tabular}{|l|c|cc|}
\hline
\multirow{2}{*}{\textbf{Domain}} & \multirow{2}{*}{\textbf{Sustainability Standard References}} & \multicolumn{2}{c|}{\textbf{Knowledge Graph Characteristics}} \\ \cline{3-4} 
                        &                                                    & \multicolumn{1}{c|}{\hspace{3ex} \#\textbf{Nodes} \hspace{4ex} }  & \#\textbf{Relations} \\ \hline
Smart Home              &  \href{https://www.canada.ca/en/health-canada/corporate/about-health-canada/reports-publications/sustainable-development/2023-2027-departmental-sustainable-development-strategy.html}{Canada Sustainable Housing}                                                  & \multicolumn{1}{c|}{69}           &   55     
\\ \hline
Healthcare              &  \href{https://www.canada.ca/en/public-health/corporate/mandate/about-agency/sustainable-development/departmental-strategy-2023-2027.html}{Canada Public Health}                                                  & \multicolumn{1}{c|}{33}           &   26   \\ \hline
E commerce              &  \href{https://www.canada.ca/en/department-finance/corporate/transparency/plans-performance/sustainable-development-strategy/2023-2027.html}{Canada Sustainable Finance}                                                  & \multicolumn{1}{c|}{33}           &   23                 \\ \hline
Transport System        &  \href{https://tc.canada.ca/en/corporate-services/transparency/sustainable-development-transport-canada/2023-2027-departmental-sustainable-development-strategy}{Canada Sustainable Transport}                                                  & \multicolumn{1}{c|}{33}           &           26         \\ \hline
\end{tabular}%
}
\end{table*}

\subsection{Qualitative Results}
This section summarizes the results obtained from each case study through the execution of various components of the \textit{SEER} framework. 
\subsubsection{SR Elicitor Component}
The \textit{SR Elicitor} component integrates SRs from the taxonomy (refer to Section \ref{sectaxo}) considering the scope of each product. The SRS for each case study and taxonomy are pre-processed and stored in vector databases. Table \ref{tabstan} shows the knowledge graph details, that is, the number of nodes and relationships defined for each standard guideline (with respect to each case study).

\begin{table*}[!h]
\caption{SR Elicitor Results}
\label{tabsrext}
\begin{tabular}{|l|l|}
\hline
\textbf{Project} & \textbf{Sustainability Concerns}                                                                                                                                                                                  \\ \hline
Smart Home       & \begin{tabular}[c]{@{}l@{}}Energy Efficiency,  Independent Living, Inclusitivity, Low cost, \\Resource Efficiency, Indoor air quality, Minimize Data Volume\end{tabular}                                        \\ \hline
Healthcare       & \begin{tabular}[c]{@{}l@{}}Trust and transparency,  Accessibility, Equality, Data Efficiency, \\Energy Consumption\end{tabular}                                                                                    \\ \hline
E commerce       & \begin{tabular}[c]{@{}l@{}}Energy Efficiency, Sustainable Data Storage  and Transmission, \\Data Management Compatibility and Accessibility, Longevity\end{tabular}                                             \\ \hline
Transport System & \begin{tabular}[c]{@{}l@{}}Energy Efficiency, Resource Management, Cost Effectiveness, \\Technical Efficiency, Inclusivity, Environmental Impact,\\ Reliability and Maintenance, Safety and Security\end{tabular} \\ \hline
\end{tabular}
\end{table*}

The SRs are then selected from the taxonomy that are relevant to each product scope. Table \ref{tabsrext} summarizes the SRs identified for the different case study. The Smart Home project demonstrates the strongest alignment with the sustainability taxonomy, integrating the largest number of SRs across environmental, social, economic, and technical domains. In the Healthcare use case, the focus is mainly on trust, equality, and accessibility, with SRs distributed across environmental, social, and technical domains. In the E-commerce application, the emphasis lies primarily on data management and energy concerns. In the Transport System, the \textit{SR Elicitor} component identifies a broader set of concerns, including energy, resource management, safety, cost, and reliability. The energy efficiency emerges as a recurring concern across all projects. The overall results indicate that the \textit{SR Elicitor} component is capable of interpreting the scope of each project and effectively deriving the key sustainability concerns relevant to their contexts.

\begin{figure*}[!b]
    \centering
    \includegraphics[width=0.6\textwidth]{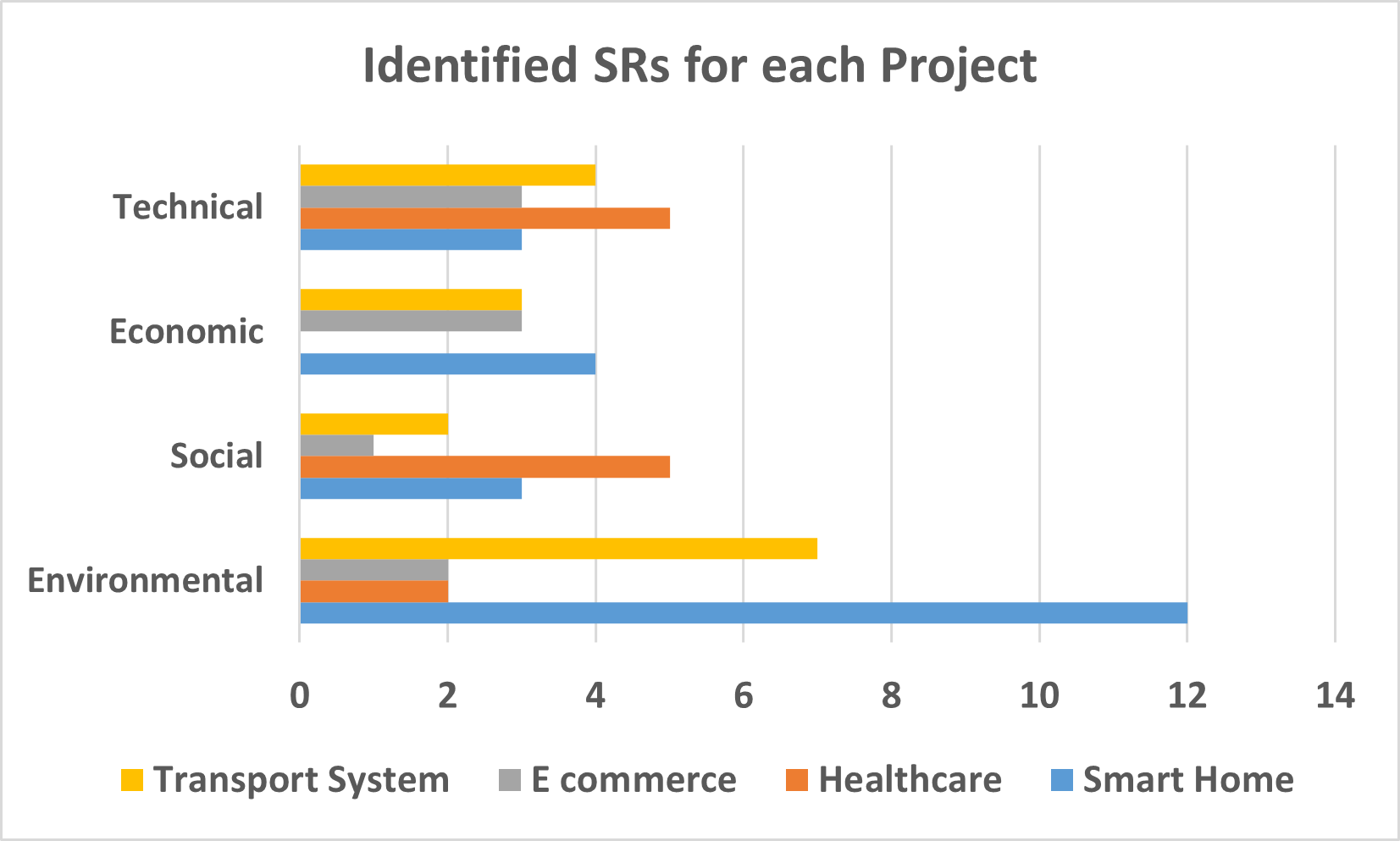}
    \caption{SRs identified in each dimension for different projects}
    \label{fig:srcount}
\end{figure*}

Figure \ref{fig:srcount} shows the exact number of SRs elicited across each dimension for different projects. In case of three projects SRs are distributed across all four dimensions. However, in healthcare project no SRs concerning economic dimension are elicited. This is due to the fact that product scope defined in the SRS does not contain treatment cost related information.

In \textit{SR Elicitor} component, the results are generated using Gemini 2.5 \cite{gemini2.5} and GPT 3.5 \cite{gpt3.5}. We have observed that both the LLMs have provided uniform responses for all four projects. The \textit{SEER} framework can be applied with any LLMs. Here, we have selected these two models due to their state-of-the-art reasoning capabilities \cite{rahman2025comparativeanalysisbaseddeepseek}.

\hspace{2ex}

\noindent\textbf{Validation}. The \textit{SR Elicitor} component is able to identify the major sustainability concerns for each project. However, it is important to assess how much the result is accurate and complete. We have (the authors) manually conducted the completeness check by reviewing the taxonomy. Relevant SRs were identified for each project based on the defined scope. This identification was done before generating experimental results to prevent bias. After the experimental process, the SRs identified by human are compared with the one identified by \textit{SR Elicitor}. Table \ref{tabval1} illustrates the results of this comparison.

\begin{table*}[!h]
\caption{Validation of SR Elicitor Results}
\label{tabval1}
\resizebox{\textwidth}{!}{%
\begin{tabular}{|c|c|c|c|c|c|}
\hline
\textbf{Project} & \textbf{\begin{tabular}[c]{@{}l@{}}\# SRs Identified \\ by Human\end{tabular}} & \textbf{\begin{tabular}[c]{@{}l@{}}\#SRs Identified \\ by SR Elicitor\end{tabular}} & \textbf{\begin{tabular}[c]{@{}l@{}}\% Intersecting\\ SRs\end{tabular}} & \textbf{\begin{tabular}[c]{@{}l@{}}\#SRs not identified \\ by SR Elicitor (false \\negatives)\end{tabular}} & \textbf{\begin{tabular}[c]{@{}l@{}}\#SRs not identified \\ by Human (false \\positives)\end{tabular}} \\ \hline
Smart Home       & 23                                                                             & 22                                                                                  & 65\%                                                                   & 8                                                                                       & 7                                                                                 \\ \hline
Healthcare       & 19                                                                             & 12                                                                                  & 63\%                                                                   & 7                                                                                       & 0                                                                                 \\ \hline
E commerce       & 10                                                                             & 9                                                                                   & 90\%                                                                   & 1                                                                                       & 0                                                                                 \\ \hline
Transport System & 17                                                                             & 16                                                                                  & 88\%                                                                   & 2                                                                                       & 1                                                                                 \\ \hline
\end{tabular}%
}
\end{table*}

The comparison shows that the SRs elicited by humans and LLM have at least more than 60\% match. For e-commerce project, the match is of the highest value 90\%. The LLM correctly identified all 9 SRs. In the smart home system, the match is around 65\%, with some SRs not being identified by LLM. Also, there are some SRs that only LLM has identified (false positives). Although in the transport system, the match value is sufficiently high, there are some SRs that are not identified by LLM (false negatives). The comparison shows that LLM-generated results are generally aligned with human judgment. However, the completeness of SR selected for a project is also dependent on the availability of SRs in the taxonomy.
\subsubsection{Relationship Integrator Component}

The \textit{Relationship Integrator} component operates in three modules. However, the first module, which involves model training, is executed only once. The remaining two modules are executed separately for each project (case study).

\begin{enumerate}
    \item In the experimental process for model training, we have fine-tuned \textit{all-MiniLM-L6-v2} $^{\ref{sentarns}}$ sentence-transformer model. The sentence transformer model \textit{all-MiniLM-L6-v2} is built for fast and efficient sentence embedding generation. It is based on \textit{Microsoft’s MiniLM-L6-H384-uncased model} and fine-tuned on over a billion sentence pairs. The model is selected due to its state-of-the-art performance in generating the fastest embedding for semantic analysis \cite{pavlyshenko}. It is fine-tuned using an aggregated source of data \cite{10.1007/978-3-030-93709-6_29}, \cite{9218190}, \cite{su16145901}. The aggregated dataset is used for both training (80\%) and validation (20\%). 
    
    The threshold for identifying related pairs was not set randomly. We have tested different similarity thresholds ranging from 0.5 to 0.9 (in steps of 0.05) on the validation dataset. The performance metrics (precision, recall, F1-score) are identified for each threshold value. Finally, we selected the threshold 0.65 that gave the best balance between precision and recall (highest F1 score) on the validation dataset. The performance of the fine-tuned model on the test dataset (for threshold 0.65) is shown in Table \ref{tab:perf}. The fine-tuned model shows a high performance on the test dataset for identifying related and unrelated requirement pairs.

    \begin{table}[!h]
    \centering
    \caption{Performance of fine-tuned model}
    \label{tab:perf}
\begin{tabular}{|c|c|c|c|c|}
\hline
\multicolumn{1}{|c|}{\textbf{Data Type}} & \multicolumn{1}{c|}{\textbf{Precision}} & \multicolumn{1}{c|}{\textbf{Recall}} & \multicolumn{1}{c|}{\textbf{Accuracy}} & \multicolumn{1}{c|}{\textbf{F1-score}} \\ \hline
Related Pairs                            & 98\%                                    & 93\%                                 & 95\%                                   & 95\%                                   \\ \hline
Unrelated Pairs                          & 92\%                                    & 97\%                                 & 95\%                                   & 95\%                                   \\ \hline
\end{tabular}
\end{table}
\item In the \textit{Related Pair Extractor} module the fine-tuned model is used for identifying related requirements for each project. The fine-tuned model embeds requirement pairs and then cosine similarity score is computed for them. Requirement pairs with cosine similarity score greater than 0.65 are kept for next step. 
Table \ref{tabrelext1} presents the results of the \textit{Related Pair Extractor} module for different projects. The second column shows the total number of possible requirement pairs by combining FR-SR, NFR-SR and SR-SR. The third column shows the actual number of requirement pairs that are semantically related.

\begin{table}[!h]
\centering
\caption{Results of Related Pair Extractor Module}
\label{tabrelext1}
\begin{tabular}{|c|c|c|}
\hline
\multicolumn{1}{|c|}{\textbf{\begin{tabular}[c]{@{}c@{}}Project \\ Domain\end{tabular}}} & \multicolumn{1}{c|}{\textbf{\begin{tabular}[c]{@{}c@{}}\#Possible \\Requirement Pairs\end{tabular}}} & \multicolumn{1}{c|}{\textbf{\begin{tabular}[c]{@{}c@{}}\#Semantically\\ Related Pairs\end{tabular}}} \\ \hline
Smart Home                                                                               & 1.219                                                                                                 & 141                                                                                                  \\ \hline
Healthcare                                                                               & 480                                                                                                  & 104                                                                                                  \\ \hline
E commerce                                                                               & 396                                                                                                  & 22                                                                                                   \\ \hline
Transport system                                                                         & 672                                                                                                  & 111                                                                                                  \\ \hline
\end{tabular}
\end{table}

\item In the final module of \textit{Relationship Integrator} we determined the exact nature of relationship among related requirement pairs. In the experimental process, we used the Gemini 2.5 \cite{gemini2.5} reasoning model to generate these results. The Gemini 2.5 models can handle long contexts and agentic workflows \cite{comanici2025gemini25pushingfrontier}. However, the approach can be replicated with any model.

Table \ref{tabrelext2} presents the results of the \textit{Relation Extractor} module. It is observed that for the first three projects, the system requirements generally satisfy the SRs by default. However, for e-commerce case study, most of the system requirements hold a negative correlation with the SRs.

\begin{table}[!h]
\caption{Relation Extraction Results}
\label{tabrelext2}
\begin{tabular}{|c|c|c|}
\hline
\textbf{Project Domain} & \textbf{\begin{tabular}[c]{@{}c@{}}\#Requirement Pairs  \\ with Positive Correlation\end{tabular}} & \textbf{\begin{tabular}[c]{@{}c@{}}\#Requirement Pairs \\ with Negative Correlation\end{tabular}} \\ \hline
Smart Home              & 119                                                                                                & 22                                                                                                \\ \hline
Healthcare              & 94                                                                                                & 10                                                                                                \\ \hline
 Transport System            & 99                                                                                                 & 12                                                                                                \\ \hline
E commerce                & 6                                                                                                  & 16                                                                                                \\ \hline
\end{tabular}
\end{table}

Figure \ref{fig:graphneg} presents a more detailed view of the number of requirements per project that show a negative correlation with SRs across various dimensions. These negative correlations are important to highlight as they are optimized in the \textit{Optimizer} component of the \textit{SEER} framework. We observed that in the smart home project, the negative correlations hold for SRs of three different dimensions. However, for other cases, healthcare and e-commerce technical and economic dimensions SRs respectively, are not satisfied mainly. This is also due to the reason that for different projects, SRs do not belong to all four dimensions. 
\begin{figure*}[!h]
    \centering
    \includegraphics[width=0.6\textwidth]{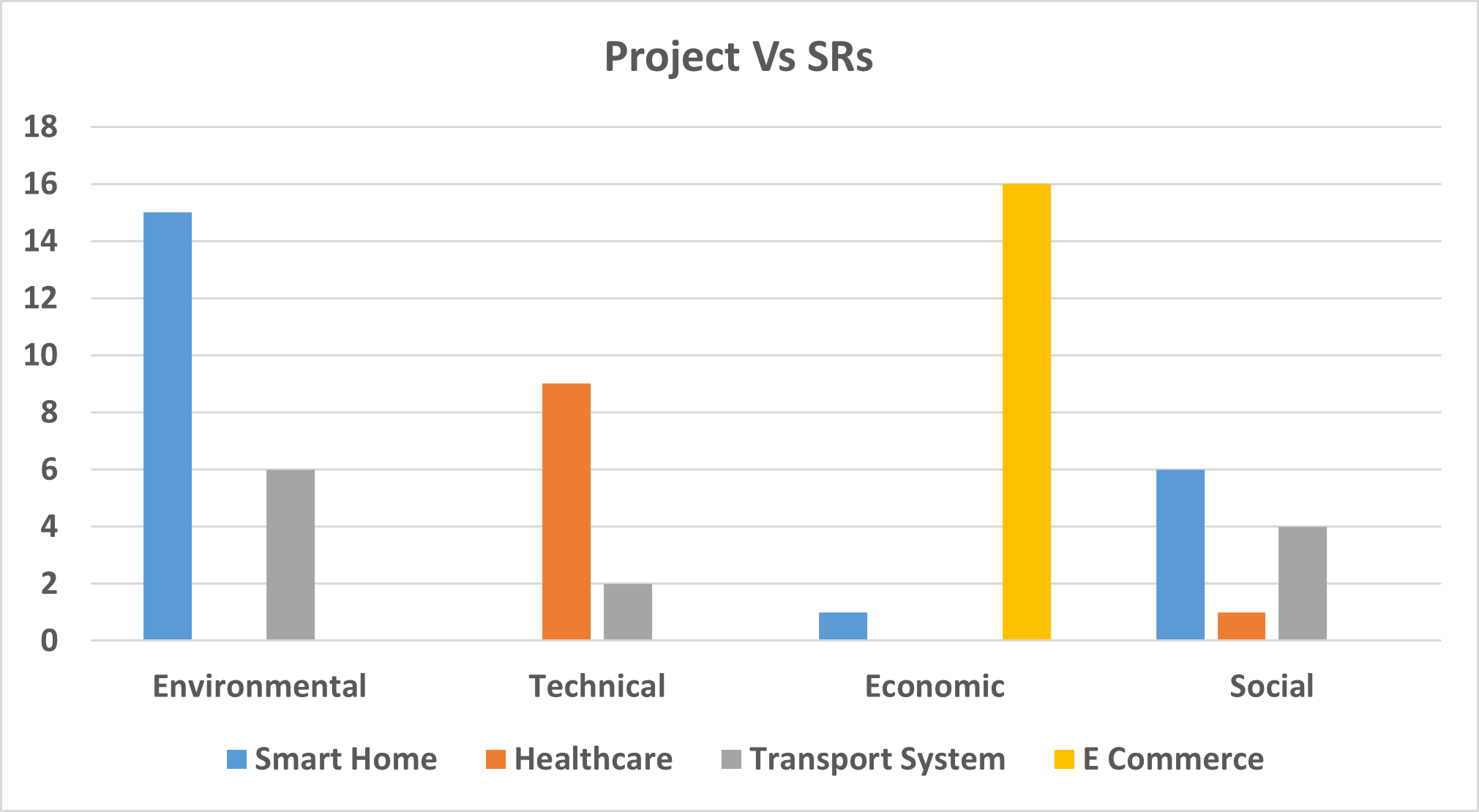}
    \caption{Negative Correlations with SRs across different projects}
    \label{fig:graphneg}
\end{figure*}
\end{enumerate}

\noindent \textbf{Validation}
The validation of the final results of \textit{Relationship Integrator} component in done in terms of consistency and trustworthiness of LLM results.

\begin{enumerate}
    \item \textit{Consistency}- The consistency is ensured by executing the RAG-based agentic framework for each project 3 consecutive times, using the same LLM (Gemini 2.5) and setting the temperature to value 0.3. The semantics of the requirements and prompt remain the same in all three executions. The positive and negative correlation results obtained in three executions are same. However, we do not ensure that tuning the prompt and requirements to a different language will guarantee exactly same result.
    \item \textit{Trustworthiness}- The trustworthiness is ensured in two steps. At first, the agent's response is checked to see if the catalogs are used for reasoning for different requirement pairs. This requires checking each response of the agent and scrutinizing the reasoning step. The results of checking are shown in Table \ref{llmre}. We find that in most cases LLM reasoning is supported by the information retrieved from the knowledge catalogs. However, in other cases the LLM agent response have shown that it did not find relevant information from the catalogs. This is consistent across all three executions.

    \begin{table*}[!h]
    \centering
    \caption{LLM Reasoning Basis}
    \label{llmre}
\begin{tabular}{|c|c|c|}
\hline
\textbf{Project} & \textbf{Catalog Referred} & \textbf{Only own reasoning} \\ \hline
Smart Home       & 99 (71\%)                       & 42                          \\ \hline
Healthcare       & 67 (66\%)                        & 37                          \\ \hline
Transport System & 87 (80\%)                       & 24                          \\ \hline
E commerce       & 11 (50\%)                       & 11                          \\ \hline
\end{tabular}
\end{table*}

    In the next step of checking trustworthiness, we have used another reasoning LLM, namely GPT 3.5\cite{gpt3.5}, to check if it generates the similar relation type for all four projects. Table \ref{relext3} shows the number of positive and negative correlations generated by GPT 3.5. Although there are some mismatches in the result but most of the requirement pairs that are labeled as negative by Gemini 2.5 were also considered as negative by GPT 3.5. In case of Smart home project the results generated by two LLMs are exactly same with a 100\% match. In heathcare project, 75\% of the positive correlations matches, and only 30\% of the negative correlation matched. However, the majority of the results demonstrate significant alignment between the outputs of the two different LLMs.

\begin{table}[!h]
\caption{Relation Extracted by GPT 3.5}
\label{relext3}
\begin{tabular}{|c|cc|cc|}
\hline
\multirow{2}{*}{\textbf{Project}} & \multicolumn{2}{c|}{\textbf{Positive Correlation}}     & \multicolumn{2}{c|}{\textbf{Negative/neutral   Correlation}} \\ \cline{2-5} 
                                  & \multicolumn{1}{c|}{\hspace{1ex} \textbf{Count}  \hspace{1ex} } & \textbf{Matched} & \multicolumn{1}{c|}{\hspace{2ex} \textbf{Count}  \hspace{4ex} }   & \textbf{Matched}     \\ \hline
\textbf{Smart Home}               & \multicolumn{1}{c|}{119}            & 100\%            & \multicolumn{1}{c|}{22}                  & 100\%                \\ \hline
\textbf{Healthcare}               & \multicolumn{1}{c|}{70}             & 75\%             & \multicolumn{1}{c|}{34}                  & 30\%                 \\ \hline
\textbf{Transport System}         & \multicolumn{1}{c|}{89}             & 90\%             & \multicolumn{1}{c|}{22}                  & 55\%                 \\ \hline
\textbf{E commerce}               & \multicolumn{1}{c|}{10}             & 60\%             & \multicolumn{1}{c|}{12}                  & 75\%                 \\ \hline
\end{tabular}
\end{table}
\end{enumerate}

\subsubsection{Optimizer Component}
In the step, we have again applied a COT prompt to an LLM (Gemini \cite{gemini2.5}) to revise system requirements (FRs or NFRs) that have negative impact on SR. A single modified requirement is proposed corresponding to each system requirement subjected to the \textit{Optimizer} component.

The results are validated in a two step process. 

\begin{enumerate}
    \item The revised requirement should not deviate from the original objective of the requirement. In real scenario, this checking should be effectively done by system analyst. However, in the experimental process, we have semantically embedded $^{\ref{sentarns}}$ the original and revised requirements and checked for their cosine similarity score. Table \ref{opres1} provides the average similarity scores for different projects. The similarity scores suggests that for smart home, healthcare and transport system project moderate amount of changes are added to generate revised requirement. However, for e commerce project minor changes are added to the original requirement. 

    \begin{table*}[!h]
    \centering
    \caption{Optimzed Requirements Seamntics}
    \label{opres1}
\begin{tabular}{|c|c|c|}
\hline
\textbf{Project} & \textbf{\begin{tabular}[c]{@{}l@{}}\#Revised\\ Requirements\end{tabular}} & \textbf{\begin{tabular}[c]{@{}l@{}}Average of Semantic\\ Similarity with Original \\Requirements\end{tabular}} \\ \hline
Smart Home       & 22                                                                       & 78.00\%                                                                                                             \\ \hline
Healthcare       & 10                                                                       & 75.00\%                                                                                                             \\ \hline
Transport System & 12                                                                       & 75.83\%                                                                                                          \\ \hline
E commerce       & 16                                                                       & 85.75\%                                                                                                          \\ \hline
\end{tabular}
\end{table*}

    \item These revised requirement pair are again provided to the \textit{Relationship Integrator} component to check relationship type. In the smart home case study, we found that out of 22 pairs, 20 were marked as having a positive correlation, while 2 were marked as neutral. In healthcare and transport system project all the revised requirements were marked with positive correlations with the associated SRs. However, for e commerce project 12 requirement pairs are marked as neutral and 4 as positive correlation. The requirement pairs marked with neutral correlation, suggests that system requirements are not negatively impacting the SRs, but still holds scope for revision. These findings are consistent. Since in previous step, we know that e-commerce requirements have only minor changes.
\end{enumerate}
All the experimental results are provided in our repository $^{\ref{github link}}$. The following section illustrates results of one of the case study. 
\subsection{Smart Home Case Study}
This section provides on overview of the results obtained for smart home case study. Let us first understand the scope of the project. An abriged version of the product scope is shown below: 

\textit{The Digital Home system, for the purposes of this document, is a system that will allow a home user to manage devices that control the environment of a home. The user communicates through a personal web page on the DigitalHome web server or on a local home server. The DH web server communicates, through a home wireless gateway device, with the sensor and controller devices in the home. }

\textit{The product is based on the Digital Home High Level Requirements Definition (HLRD 2010) is intended as a prototype, which will allow business decisions to be made about future development of a commercial product. The scope of the project will be limited to the management of devices which control temperature, humidity, security, and power to small appliances and lighting units, through the use of a web-ready device. 
} 

The product scope is obtained from the software requirement specification document of a digital home system from PURE \cite{pure} dataset.

The outcomes of each component for this case study are as follows-
\begin{enumerate}
    \item \textit{SR Elicitor}- The Canada Standard Guidelines for housing (refer to Table \ref{tabstan}) are referred while eliciting SRs for this project. Some of the important SRs concerning energy efficiency, affordability, independent living and compatibility elicited from the taxonomy are as follows-
\begin{itemize}
    \item  SR$_1$: The system should be a low-cost solution for home automation. \\ Dimension:	Economic, Category: Affordability
    \item  SR$_2$: Smart home systems should support independent living for residents. \\ Dimension:	Social, Category:	Autonomy
    \item  SR$_3$: The system must establish standardized protocols and ensure compatibility between existing IoT 
devices and platforms. \\ Dimension:	Technical, Category:	Compatibility
\item  SR$_4$: The system must monitor and control lighting levels to reduce unnecessary energy usage. \\ Dimension: Environmental, Category:	Energy efficiency
\end{itemize}

The complete list of elicited SRs (22 in number) are given in github repository$^{\ref{github link}}$. 

\item \textit{Relationship Integrator}- This component shows the relationship among the elicited SRs with the system requirements. Table \ref{casres} provides a snippet of the results obtained. $FR30$ holds a positive relation with $SR4$ as it directly supports energy management through application monitoring. However, $FR31$ that requires managing many power switches negatively influences $SR4$. 

\begin{table*}[!h]
\caption{Relation Extraction for Smart Home}
\label{casres}
\resizebox{\textwidth}{!}{%
\begin{tabular}{|l|l|l|l|}
\hline
\multicolumn{1}{|c|}{\textbf{System Requirement (FR/NFR)}}                                                                                                        & \multicolumn{1}{c|}{\textbf{SR}}                                                                                                           & \multicolumn{1}{c|}{\textbf{Correlation}} & \multicolumn{1}{c|}{\textbf{Reasoning (abriged)}}                                                                                       \\ \hline
\begin{tabular}[c]{@{}l@{}}FR30: Appliance Manager shall provide\\ management of  small appliances,\\ including lighting units.\end{tabular}                        & \begin{tabular}[c]{@{}l@{}}SR4: The system must monitor\\ and  control lighting levels to\\ reduce  unnecessary energy usage.\end{tabular} & Positive                                  & \begin{tabular}[c]{@{}l@{}} Appliance control supports reduction\\ of wasted energy. \end{tabular}                                                                                  \\ \hline
\begin{tabular}[c]{@{}l@{}}FR9: User can set thermostat temperatures \\ (60–80 °F, 1° increments).\end{tabular}                                                   & \begin{tabular}[c]{@{}l@{}}SR4: Reduce unnecessary energy\\ usage.\end{tabular}                                                                                                   & Positive                                  & \begin{tabular}[c]{@{}l@{}}Fine-grained control allows optimization of\\  heating/cooling, reducing energy waste.\end{tabular}          \\ \hline
\begin{tabular}[c]{@{}l@{}}NFR4: The system must be highly reliable \\ ($\leq$1 failure per 10,000 hrs).\end{tabular}                                                  & \begin{tabular}[c]{@{}l@{}}SR2: Support independent living\\ for  residents.\end{tabular}                                                  & Positive                                  & \begin{tabular}[c]{@{}l@{}}Reliability ensures safety and autonomy\\ for residents,  especially elderly or disabled.\end{tabular}       \\ \hline
\begin{tabular}[c]{@{}l@{}}FR35: DigitalHome Planner allows preset \\home  parameters (temperature, humidity,\\ appliances, etc.)  for time periods.\end{tabular} & \begin{tabular}[c]{@{}l@{}}SR4: Reduce unnecessary energy\\ usage.\end{tabular}                                                                                                      & Positive                                  & \begin{tabular}[c]{@{}l@{}}Scheduling automation avoids energy waste \\ when spaces are unoccupied.\end{tabular}                        \\ \hline
\begin{tabular}[c]{@{}l@{}}NFR11: Development shall use IEEE and\\ HO4710  standards, with proper documentation\\ and modularity.\end{tabular}                      & \begin{tabular}[c]{@{}l@{}}SR3: Establish standardized\\ protocols and  ensure compatibility\\ between IoT devices.\end{tabular}             & Positive                                  & \begin{tabular}[c]{@{}l@{}}Use of international standards directly\\ fosters  device interoperability.\end{tabular}                     \\ \hline
\begin{tabular}[c]{@{}l@{}}FR31: Appliance Manager shall manage up to\\ 100  power switches (115V, 10A).\end{tabular}                                             & \begin{tabular}[c]{@{}l@{}}SR4: Reduce unnecessary energy\\ usage.\end{tabular}                                                                                             & Negative                                  & \begin{tabular}[c]{@{}l@{}}Supporting many high-power switches \\could lead to  excessive energy consumption\\ if unmanaged.\end{tabular} \\ \hline
\begin{tabular}[c]{@{}l@{}}FR17: The system shall be compatible with\\ centralized  HVAC systems (gas, oil, electricity, \\solar, etc.).\end{tabular}               & \begin{tabular}[c]{@{}l@{}}SR1: Low-cost solution for home\\ automation \end{tabular}.                                                                                                & Negative                                  & \begin{tabular}[c]{@{}l@{}}Broad HVAC compatibility raises\\ complexity and  costs, reducing affordability.\end{tabular}                \\ \hline
\begin{tabular}[c]{@{}l@{}}NFR1: Displays of environmental conditions\\ shall be  updated at least every 2 seconds.\end{tabular}                                  & SR4: Energy efficiency.                                                                                                                    & Negative                                  & \begin{tabular}[c]{@{}l@{}}Frequent updates increase device power\\ usage and  network load, lowering energy \\ efficiency.\end{tabular}   \\ \hline
\end{tabular}%
}
\end{table*}

\item \textit{Optimizer}- In this step, we revise the three requirements that show a negative correlation with $SR1$ and $SR4$. Table \ref{casres2} shows the original, revised requirement and their similarity scores. It can be observed each revised requirement holds original intention while adding constraints for SR satisfaction.

\begin{table*}[!h]
\caption{Requirement Optimization for Smart Home}
\label{casres2}
\resizebox{\textwidth}{!}{%
\begin{tabular}{|l|l|l|}
\hline
\multicolumn{1}{|c|}{\textbf{Original Requirement}}                                                                                                                                                                & \multicolumn{1}{c|}{\textbf{Revised Requirement}}                                                                                                                                                                                                                                                                                          & \multicolumn{1}{c|}{\textbf{Similarity Score}} \\ \hline
\begin{tabular}[c]{@{}l@{}}FR 31: The Appliance Manager shall be able to \\ manage up to one hundred 115 volt, 10 amp \\ power switches.\end{tabular}                                                              & \begin{tabular}[c]{@{}l@{}}The Appliance Manager shall be able to manage \\ up to one hundred 115 volt, 10 amp power switches,\\  while supporting automated energy optimization features\\  (e.g., standby power reduction, usage scheduling, \\ or priority-based switching) to minimize unnecessary energy \\ consumption.\end{tabular} & \textbf{0.53}                                  \\ \hline
\begin{tabular}[c]{@{}l@{}}FR17: The system shall be compatible with \\ centralized HVAC systems (gas, oil, electricity, \\ solar, or a combination). The system shall adhere to \\ ASHRAE standards.\end{tabular} & \begin{tabular}[c]{@{}l@{}}The system shall be compatible with at least one standard \\ centralized HVAC system (gas, oil, electricity, or solar), \\ with modular expandability to support additional types as \\ needed, while maintaining compliance with ASHRAE standards.\end{tabular}                                                & \textbf{0.55}                                  \\ \hline
\begin{tabular}[c]{@{}l@{}}NFR1: Displays of environmental conditions\\  shall be updated at least every two seconds.\end{tabular}                                                                                 & \begin{tabular}[c]{@{}l@{}}Displays of environmental conditions shall be updated \\ at least every two seconds when real-time monitoring is \\ required, and shall otherwise support adaptive update \\ intervals (e.g., event-driven or user-configurable) to \\ optimize energy efficiency.\end{tabular}                                 & \textbf{0.51}                                  \\ \hline
\end{tabular}%
}
\end{table*}
\end{enumerate}

\subsection{Discussion}
The experimental results from four diverse case studies demonstrate the effectiveness of the \textit{SEER} framework in eliciting SRs and in analyzing and optimizing both FRs and NFRs with consideration of those SRs. However, there are some threats to validity of the experimental results that needs discussion-

\begin{enumerate}
    \item The first concern is regarding human annotation of SRs for validating results of \textit{SR Elicitor}. Although the authors have experience in working with sustainability project and software requirements, it does not ensure complete correctness and accuracy. The major limitation of this process is related to human attention span. The sustainability taxonomy contains a significant number of SRs, it may be possible that some SRs are incorrectly identified and some are missed. In contrast, the experiments conducted using two LLMs generated consistent SRs across all four projects. This consistency enhances the reliability of the \textit{SEER} framework, suggesting that human validation can serve as a supporting measure rather than a primary determinant.
    \item The second concern is related to the elicitation of sustainability concerns across all four dimensions. In the experimental result, we found that SRs for all projects do not include all four dimension. There are two possible reasons for such scenarios.
    \begin{itemize}
        \item The SRs in the taxonomy are limited. The generated results can be more uniform when the referred SR taxonomy is more inclusive for all domains.
        \item The SR elicitation not only depends on the taxonomy, but also on the completeness of the SRS documents. When the SRS document precisely captures the scope of the product, results of SR elicitation will be much better.
    \end{itemize}
    \item The \textit{SEER} framework intensely depends on LLM for result generation across all three components. This makes LLM stochasticity \cite{klishevich2025measuringdeterminismlargelanguage} an important concern. However, we have applied few preventive measures \cite{klishevich2025measuringdeterminismlargelanguage} to handle uncertainity- (i) setting low temperature value (ii) a guided structured prompt (iii) few-shot examples (iv) retrieval augmented response.

    The choice of agentic-RAG approach instead of normal RAG also ensures better reasoning \cite{singh2025agenticretrievalaugmentedgenerationsurvey}. The agent can perform multi-step reasoning using information from the knowledge bases, before producing the final output. 
    \item The sensitivity of LLM response for different prompts could be an issue. However, in this work we have defined structured prompts $^{\ref{claude}}$ and used it consistently for all projects in the experiments. The prompts are made open-sourced as it is important for the reproducibility of the \textit{SEER} framework.
    \item The experimental process is conducted to only four domains. This does not necessarily generalize the effectiveness or applicability for other domains like- agriculture, energy sector. However, \textit{SEER} framework has shown satisfactory results for four diverse domains.
\end{enumerate}

\section{Conclusion}\label{con}
Sustainable software engineering has gained significant traction in recent years as awareness of the environmental, social, and economic impacts of software systems continues to grow. Despite of this progress, there remains a notable lack of a comprehensive framework (tool support) that enables software developers to incorporate sustainability considerations right from the initial requirements engineering phase. The \textit{SEER} framework addresses this gap by offering an automated approach that integrates sustainability concerns directly into the requirements engineering process. The framework evaluates and optimizes system requirements to ensure alignment with defined sustainability goals.

The effectiveness of the \textit{SEER} framework has been demonstrated through extensive experimentation across multiple software projects of diverse domains (housing, healthcare, transport, e-commerce). These experiments have primarily leveraged the Gemini 2.5 reasoning model to perform agent-based analysis and optimization. However, the design of \textit{SEER} is flexible, allowing it to be adopted with any LLM capable of supporting agent-based reasoning.

\textit{SEER} framework focuses on four core dimensions of sustainability- environmental, technical, social, and economic. A potential extension of \textit{SEER} could incorporate a fifth dimension, referred to as the \textit{individual} dimension, which would involve evaluating the values, beliefs, and preferences of end-users. Integrating this dimension presents additional challenges, as it requires a nuanced understanding of human factors and user perspectives.

\section*{Acknowledgements} \noindent This work takes place within the framework of the DoE 2023-2027 (MUR, AIS.DIP. ECCELLENZA2023\_27.FF project) and is partially funded by the EcoDigify Erasmus+ project co-funded by the European Union (Project Reference: 2024-1-SE01-KA220-HED-000250071, CUP: H74E24000050006).

\bibliographystyle{elsarticle-num}
\bibliography{Reference}
\section*{Appendix}
\subsection*{Knowledge Graph Ontology}
\begin{figure*}[!h]
        \centering
    \includegraphics[width=0.5\textwidth, frame]{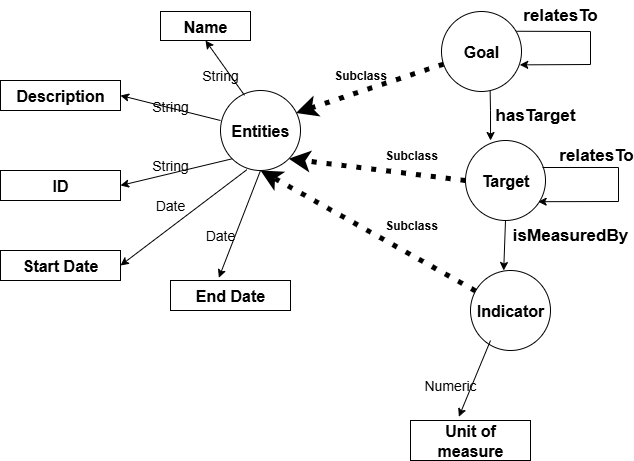}
        \caption{Knowledge Graph Ontology}
        \label{fig:onto}
    \end{figure*}

A diagrammatic representation of the ontology is shown in Figure \ref{fig:onto}. The description of the ontology are as follows-
    \begin{enumerate}
        \item The ontology consists of the following classes-
        \begin{itemize}
            \item \textit{Goals}- A high-level sustainability objective (e.g., "Reduce greenhouse gas emissions"). The goals in the standard documents are usually defined considering the UNs SDG.
            \item \textit{Targets}- A specific, measurable objective that contributes to a Goal (e.g., "Achieve 30\% reduction in emissions by 2030").
            \item \textit{Indicators}- A metric used to track progress towards a Target (e.g., "CO2 equivalent emissions (tonnes)").
        \end{itemize}
        The attributes associated with these classes are: name, description, ID, startDate, endDate, unitOfMeasure (for indicator only)
        \item The ontology consists of the following relationships-
        \begin{itemize}
            \item \textit{hasTarget}: Connects a Goal to a Target as defined specifically in the document (as shown in  Figure \ref{fig:standard}).
\item \textit{isMeasuredBy}: Connects a Target to an Indicator as defined specifically in the document (as shown in Figure \ref{fig:standard}).
\item \textit{relatesTo}: This relationship is defined for connections between Goals or Targets that might influence each other without a direct hierarchical link (e.g., one goal supporting another, or one target influencing another).

        \end{itemize}
    \end{enumerate}
    
\subsection*{LLM Prompt for Knowledge Graph Creation}

\textbf{Instruction Prompt}:

"""\textit{You are an AI assistant that extracts sustainability knowledge from a given standard document as input into a knowledge graph.}

[Input Document]

\textit{You must follow this ontology for constructing the knowledge graph:}

\textit{Classes:}

1. Goal: A high-level sustainability objective. Attributes: id, name, description, startDate, endDate

2. Target: A specific, measurable objective that contributes to a Goal. Attributes: id, name, description, startDate, endDate

3. Indicator: A metric used to track progress towards a Target. Attributes: id, name, description, startDate, endDate, unitOfMeasure

\textit{Relationships (use these exact property names in output)}:

1. hasTarget (Goal → Target): Connects a Goal to its specific Targets.

2. isMeasuredBy (Target → Indicator): Connects a Target to Indicators that measure progress.

3. relatesTo : Generic relation to capture links between Goals, Targets, or Indicators. Use when one entity influences, supports, or is otherwise related to another

\textit{Task: From the input text, identify Goals, Targets, and Indicators and output a structured JSON-LD array. Follow these output rules strictly:}

\textit{1. Every Goal/Target/Indicator object must include @id, @type, and ex:name.}

\textit{2. Use ex:hasTarget, ex:isMeasuredBy, and ex:relatesto to express relationships. ex:relatesto values must be one @id or an array of @ids.}

\textit{3. Use null for unknown attributes (e.g., startDate/endDate).}

\textit{4. When multiple indicators measure a target, ex:isMeasuredBy should be an array of {"@id": "..."}.}

\textit{5. Keep output only the JSON-LD array (no extra text).}

\textit{Example Input Text:}

Goal G1: Ensure access to affordable, safe, and sustainable housing for all.

Target: Reduce chronic homelessness by at least 31

Indicator: Number of chronically homeless individuals.

Goal G2: Make cities inclusive, safe, resilient, and sustainable.

Target: By 2030, ensure access for all to adequate housing and basic services.

Indicator: Proportion of urban population living in slums.

Goal G3: Encourage inclusive and sustainable economic growth in Canada.

Target: By 2030, achieve full and productive employment and decent work for all.

Indicator: Unemployment rate (\% of population).

\textit{Output:}

\{
  "entities": [
    \{
      "id": "G1",
      
      "type": "Goal",
      
      "name": "Ensure access to affordable, safe, and sustainable housing for all",
      
      "description": "A sustainability goal focused on improving housing accessibility and affordability."
    \},
    \{
      "id": "T1",
      
      "type": "Target",
      
      "name": "Reduce chronic homelessness by at least 31\% by March 2024",
      
      "description": "A measurable target contributing to affordable housing."
    \},
    \{
      "id": "I1",
      
      "type": "Indicator",
      
      "name": "Number of chronically homeless individuals",
      
      "unitOfMeasure": "count"
    \},
    \{
      "id": "G2",
      
      "type": "Goal",
      
      "name": "Make cities inclusive, safe, resilient, and sustainable",
      
      "description": "A sustainability goal aligned with SDG 11."
    \},
    \{
      "id": "T2",
      
      "type": "Target",
      
      "name": "By 2030, ensure access for all to adequate housing and basic services",
      
      "description": "A target contributing to urban sustainability."
    \},
    \{
      "id": "I2",
      
      "type": "Indicator",
      
      "name": "Proportion of urban population living in slums",
      
      "unitOfMeasure": "percentage"
    \},
    \{
      "id": "G3",
      
      "type": "Goal",
      
      "name": "Encourage inclusive and sustainable economic growth in Canada",
      
      "description": "A goal focused on fostering economic sustainability."
    \},
    \{
      "id": "T3",
      
      "type": "Target",
      
      "name": "By 2030, achieve full and productive employment and decent work for all",
      
      "description": "A target contributing to sustainable economic growth."
    \},
    \{
      "id": "I3",
      
      "type": "Indicator",
      
      "name": "Unemployment rate",
      
      "unitOfMeasure": "percentage"
    \}
  ],
  "relationships": [
    \{"from": "G1", "to": "T1", "type": "hasTarget"\},
    
    \{"from": "T1", "to": "I1", "type": "isMeasuredBy"\},
    
    \{"from": "G2", "to": "T2", "type": "hasTarget"\},
    
    \{"from": "T2", "to": "I2", "type": "isMeasuredBy"\},
    
    \{"from": "G3", "to": "T3", "type": "hasTarget"\},
    
    \{"from": "T3", "to": "I3", "type": "isMeasuredBy"\},
    
    \{"from": "G1", "to": "G2", "type": "relatesTo"\},
    
    \{"from": "G1", "to": "G3", "type": "relatesTo"\},
    
    \{"from": "G2", "to": "G3", "type": "relatesTo"\}
  ]
\}

]

"""

\subsection*{LLM Agent Prompt for Knowledge Extraction}
\textbf{Iterative Instruction Prompt}:

"""\textit{You are an intelligent agent with access to a sustainability knowledge graph (KG).
Your task is to analyze a product chunk and identify related sustainability goals, targets, and indicators.}

\textit{You have access to following tools:}

{tools}

\textit{Instructions:}

1. Identify key concepts/entities in the chunk.

2. Query the KG (internally) to find related sustainability goals and targets.

3. Reason step-by-step (chain-of-thought) why each goal is relevant.

4. Store a memory entry with the following format:

Memory Entry:

Chunk ID: {chunk_id}

Key Concepts: <list>

Related entities: <list>

\textit{Use the following format for reasoning:}

    \textit{Question: the input question you must answer}
    
    \textit{Thought: you should always think about what to do
    Action: the action to take, should be one of [{tool_names}]}
    
    \textit{Action Input: the input to the action}
    
    \textit{Observation: the result of the action
    ... (this Thought/Action/Action Input/Observation can repeat N times)}
    
    \textit{Thought: I now know the final answer}
    
    \textit{Final Answer: the final answer to the original input question}

    \textit{Begin!}

    Question: {input}
    Thought:{agent_scratchpad}

"""

\textbf{Final Syntheis Prompt}:

"""\textit{You have processed multiple product chunks and stored intermediate memory entries.}

\textit{Memory Entries:}

{memory_entries}

\textit{Task:}

Provide a final synthesis of all relevant sustainability goals.

For each goal, include:

- Related targets/indicators

- Interdependencies with other goals

Output in the following structured format:

Final Sustainability Goal Analysis:

Goal : <Goal Name>

  Related Targets/Indicators: <list>
  
  Interdependencies: <list>

  \textit{Use the following format for reasoning:}

    \textit{Question: the input question you must answer}
    
    \textit{Thought: you should always think about what to do
    Action: the action to take, should be one of [{tool_names}]}
    
    \textit{Action Input: the input to the action}
    
    \textit{Observation: the result of the action
    ... (this Thought/Action/Action Input/Observation can repeat N times)}
    
    \textit{Thought: I now know the final answer}
    
    \textit{Final Answer: the final answer to the original input question}

    \textit{Begin!}

    Question: {input}
    Thought:{agent_scratchpad}

"""

\subsection*{Prompt for Mixed Agent}

\textbf{Instruction Prompt}:

"""\textit{You are an AI assistant for requirements analysis, You will help to identify potential conflicts or positive dependencies among the following  pairs of requirements specifications.}

\textit{ You have access to the following tool:}

{tools}

The tools return following relevant information-

1.  Dependency category information among requirements

2. Conflict information about different sustainability category. If sustainability category does not match the query generate result based on your own input.

3. Conflict information about different NFR category. If NFR category does not match the query generate result based on your own input.

You may use one or more tools for generating your answer.

Here are some examples of correlations (positive or negative) among functional and sustainability requirements-

FR1: The system shall optimize server workload distribution to improve performance.

SR1: The system should minimize energy consumption of computing resources.

Correlation: Positive – load balancing improves energy efficiency, supporting sustainability.

FR2: The DigitalHome shall allow monitoring of electricity consumption of each device.

SR2: The system should promote energy savings by providing users feedback on consumption.

Correlation: Positive – monitoring enables user awareness and energy-saving actions.

FR3: The system shall enable remote access to home appliances.

SR3: The system should reduce unnecessary travel for household management.

Correlation: Positive – remote control avoids physical travel, reducing carbon footprint.

FR4: The system shall store high-resolution logs of all user interactions for 2 years.

SR4: The system should minimize data storage and reduce resource usage.

Correlation: Negative – long-term data storage conflicts with resource efficiency.

FR5: The system shall provide real-time video surveillance accessible from anywhere.

SR5: The system should minimize network bandwidth consumption.

Correlation: Negative – real-time video streaming consumes significant bandwidth.

FR6: The application shall provide instant AI recommendations by continuously running inference models in the background.

SR6: The system should reduce energy usage of computational processes.

Correlation: Negative – continuous inference increases energy consumption.

NFR1: The system shall achieve 99\% uptime availability.

SR1: The system should ensure reliable access to services to reduce resource waste from repeated retries.

Correlation: Positive – higher availability reduces redundant retries and wasted energy.

NFR2: The application shall respond to user requests within 2 seconds.

SR2: The system should optimize processing to minimize CPU cycles and energy use.

Correlation: Positive – faster response times can be achieved via optimized algorithms that also save energy.

NFR3: The system shall support modular architecture.

SR3: The system should extend product lifecycle by allowing upgrades instead of replacements.

Correlation: Positive – modularity enables sustainable upgrades, reducing e-waste.

NFR4: The system shall maintain logs with detailed debug information at all times.

SR4: The system should minimize storage requirements and energy used for data management.

Correlation: Negative – continuous logging increases storage and energy demands.

NFR5: The system shall support ultra-high availability with full data replication across five regions.

SR5: The system should minimize infrastructure footprint and energy use.

Correlation: Negative – more replication for uptime increases infrastructure and carbon emissions.

NFR6: The system shall provide 4K video streaming quality by default.

SR6: The system should minimize network bandwidth consumption.

Correlation: Negative – default high-resolution streaming increases bandwidth and energy usage.

NFR7: The system shall encrypt all data in transit and at rest.

SR7: The system should minimize computational overhead and energy usage.

Correlation: Mixed – encryption improves security but increases CPU load; sustainable if optimized ciphers are used.

NFR8: The system shall automatically scale resources based on user demand.

SR8: The system should reduce idle resource consumption.

Correlation: Positive when scaling down resources during low usage; negative if scaling up excessively without limits.

Use the following format:

\textit{Question: the input question you must answer}

\textit{Thought: you should always think about what to do}

\textit{Action: the action to take, should be one of [{tool_names}]}

\textit{Action Input: the input to the action}

\textit{Observation: the result of the action
... (this Thought/Action/Action Input/Observation can repeat N times)}

\textit{Thought: I now know the final answer}

\textit{Final Answer: the final answer to the original input question}

The output format should be

Requirement 1:

Requirement 2:

Relation Type:

Reason:

Begin!

Question: {input}

Thought:{agent_scratchpad}
"""

\subsection*{Prompt for Sustainability Agent}

\textbf{Instruction Prompt}

"""
\textit{You are an AI assistant for requirements analysis, You will help to identify potential conflicts or positive dependencies among the following  pairs of sustainability requirements specifications.}

\textit{ You have access to the following tool:}

{tools}

The tool return the following relevant information-

1. Conflict information about different sustainability category. If sustainability category does not match the query generate result based on your own input.

Here are some examples of correlations (positive or negative) among sustainability requirements-

SR1 (Energy Efficiency): The system should minimize energy consumption of computing resources.

SR2 (Cost Efficiency): The system should reduce operational costs by optimizing infrastructure usage.

Correlation: Positive – lowering energy use also reduces electricity costs.

SR3 (Accessibility): The system should provide services accessible via low-bandwidth networks.

SR4 (Inclusiveness): The system should ensure equal access in developing regions with limited infrastructure.

Correlation: Positive – supporting low-bandwidth directly improves accessibility in underserved regions.

SR5 (Data Retention): The system should store all user data for transparency and accountability.

SR6 (Resource Efficiency): The system should minimize storage usage and reduce data center energy demand.

Correlation: Negative – long-term data retention conflicts with minimizing storage and energy consumption.

SR7 (High Availability): The system should ensure 99.99\% uptime with global server replication.

SR8 (Environmental Impact): The system should reduce carbon footprint by minimizing hardware and infrastructure.

Correlation: Negative – more replication for uptime increases infrastructure and carbon emissions.

Use the following format:

\textit{Question: the input question you must answer}

\textit{Thought: you should always think about what to do
Action: the action to take, should be one of [{tool_names}]}

\textit{Action Input: the input to the action}

\textit{Observation: the result of the action
... (this Thought/Action/Action Input/Observation can repeat N times)}

\textit{Thought: I now know the final answer}

\textit{Final Answer: the final answer to the original input question}

The output format should be

Requirement 1:

Requirement 2:

Relation Type:

Reason:

Begin!

Question: {input}

Thought:{agent_scratchpad}

\subsection*{LLM Prompt for Optimizer Component}

\textit{Instruction Prompt}

"""\textit{You are an expert requirements engineer and sustainability analyst.}

\textit{You will be given:}

- A requirement A (Functional Requirement — FR — or Non-Functional Requirement — NFR).

- A Sustainability Requirement B (SR).

Situation: Requirement A as written negatively impacts SR B.

Your goal: Revise requirement A so the negative impact on SR B is removed or substantially reduced, while preserving the original meaning, objective, and acceptance criteria of A.

Use the following format:

1. \textit{Parse and classify A (FR or NFR). Identify core intent and success criteria.}

2. \textit{Parse SR: identify sustainability objectives B}.

3. \textit{Identify nature of conflict between A and B.}

4. \textit{Generate 3 candidate revisions:}

   - \textit{Minimal tweak (light modification)}
   
   - \textit{Moderate change (introduce constraints/optimizations)}
   
   - \textit{Alternative approach (different design achieving same objective)}
   
5. \textit{For each: derive measurable acceptance criteria + risks.}

6. \textit{Score each candidate on preservation, SR impact reduction, confidence.}

7. \textit{Select recommended candidate + justify}.

Output in following JSON format:

\{
  "original_requirement": "<text>",
  
  "requirement_type": "FR|NFR",
  
  "sustainability_requirement": "<text>",
  
  "candidates": [
   
    \{
      "label": "Minimal",
      
      "revised_requirement": "<text>",
      
      "preservation_score": <0-100>,
      
      "estimated_SR_impact_reduction": "<Low|Medium|High or \%>",
      
      "confidence": "<Low|Medium|High>",
      
      "acceptance_criteria": ["criterion 1", "criterion 2"],
      
      "residual_risks": ["risk 1"]
    \},
    
    \{ "label": "Moderate", ... \},
    
    \{ "label": "Alternative", ... \}
  ],
  
  "recommended": \{
  
    "revised_requirement": "<text>",
    
    "justification": "<brief>"
  
  \},

\}

"""
\end{document}